\documentclass{IEEEtaes}
\IEEEoverridecommandlockouts
\usepackage{cite}
\usepackage{amsmath,xfrac,amssymb,amsfonts,bm,cancel,mathtools}
\usepackage{algorithm}\usepackage[noend]{algpseudocode}
\usepackage{graphicx}
\usepackage{textcomp}
\usepackage{xcolor,booktabs,wrapfig,subfig}
\usepackage{verbatim} 
\def\BibTeX{{\rm B\kern-.05em{\sc i\kern-.025em b}\kern-.08em
    T\kern-.1667em\lower.7ex\hbox{E}\kern-.125emX}}

\usepackage[letterspace=-90]{microtype} 
\usepackage{soul} 

\usepackage{mdframed}    

\renewcommand{\thesection}{\Roman{section}}

\usepackage{amsthm}
\newtheoremstyle{mytheoremstyle} 
{\topsep}                    
{\topsep}                    
{}                           
{}                           
{\scshape}                        
{}                           
{.5em}                       
{\textcolor{black}{\thmname{#1}\thmnumber{ #2}\thmnote{ (#3):}}}  

\theoremstyle{mytheoremstyle}

\newtheorem{axiom}{Axiom}
\newtheorem{definition}{Definition}
\newtheorem{lemma}{Lemma}
\newtheorem{proposition}{Proposition}
\newtheorem{corollary}{Corollary}
\newtheorem{observation}{Observation}

\newcommand{\e}{\mathrm{e}}
\newcommand{\projFcn}{\mathcal{S}}
\newcommand{\spaceOfMarks}{\mathcal{M}}
\newcommand{\likelihood}{L}

\newcommand{\RFSXi}{\Xi}
\newcommand{\RFSTheta}{\Theta}

\newcommand{\groundTruth}{GT}
\newcommand{\probabilityOfDetection}{PoD}
\newcommand{\expectedProbabilityOfDetection}{E\probabilityOfDetection{}}
\newcommand{\locallyCompactSecondCountableHausdorff}{LCSCH}
\newcommand{\completelySeparableMetric}{CSM}

\newcommand{\StateSet}{X}

\newcommand{\ObjectGenRFS}{D}
\newcommand{\ClutterRFS}{C}

\newcommand{\xThreeDee}{x}
\newcommand{\xVelocityThreeDee}{\dot{x}}
\newcommand{\yThreeDee}{y}
\newcommand{\yVelocityThreeDee}{\dot{y}}
\newcommand{\zThreeDee}{z}
\newcommand{\zVelocityThreeDee}{\dot{z}}
\newcommand{\widthThreeDee}{\omega}
\newcommand{\heightThreeDee}{h}

\newcommand{\Expect}[1]{\mathrm{E}\left[\ #1 \ \right]}

\newcommand{\X}{\mathcal{X}}
\newcommand{\Z}{\mathcal{Z}}

\renewcommand{\d}{\mathrm{d}}
\newcommand{\T}{^\top}

\begin{document}

\title{Occlusion-Aware Multi-Object Tracking via Expected Probability of Detection}


 \author{Jan Krejčí}
\affil{University of West Bohemia, Pilsen, Czech Republic}

\author{Oliver Kost}
\affil{University of West Bohemia, Pilsen, Czech Republic}

\author{Yuxuan Xia}
\affil{Shanghai Jiaotong University, Shanghai, China}

\author{Lennart Svensson}
\member{Senior Member, IEEE}
\affil{Chalmers University of Technology, Göteborg, Sweden}

\author{Ondřej Straka}
\member{Member, IEEE}
\affil{University of West Bohemia, Pilsen, Czech Republic}
\receiveddate{This research was partially supported by the European Union under the project ROBOPROX (reg. no. CZ.02.01.01/00/22\_008/0004590) and by the Czech Science Foundation (GACR) under grant GA 25-16919J.}
\maketitle

\begin{abstract}
This paper addresses multi-object systems, where objects may occlude one another relative to the sensor.
The standard point-object model for detection-based sensors is enhanced so that the probability of detection considers the presence of all objects. A principled tracking method is derived, assigning each object an expected probability of detection, where the expectation is taken over the reduced Palm density, which means conditionally on the object's existence. The assigned probability thus considers the object's visibility relative to the sensor, under the presence of other objects. Unlike existing methods, the proposed method systematically accounts for uncertainties related to all objects in a clear and manageable way. The method is demonstrated through a visual tracking application using the multi-Bernoulli mixture (MBM) filter with marks.
\end{abstract}


\section{Introduction}
Multi-object tracking (MOT) refers to estimating the number and locations of multiple moving objects, given sets of noisy detections~\cite{MTT-Survey:2015}.
The corresponding tracking algorithms are key, e.g., in aerial and naval security~\cite{BlackmanPopoli:1999} or autonomous driving~\cite{KITTI-dataset:2012}. 
Most sensors require a clear \emph{line of sight} to detect objects, but \emph{occluding} moving objects can block this visibility.
When a tracking algorithm fails to predict occlusion, it may lose sight of occluded objects, which can hinder users in safety-critical situations.

Numerous methods have been proposed to handle such occlusion situations across diverse domains and sensors.
The strategies for handling occlusion can vary significantly, ranging from solutions based on \emph{ad-hoc} approaches to those grounded in \emph{physical models}. 
In particular, the methods can be classified into \emph{(i)} fully ad-hoc, \emph{(ii)} combined ad-hoc and model-based, and \emph{(iii)} fully model-based strategies.
Note that by an \emph{ad-hoc} solution, we mean that the solution includes some procedure that is \emph{inconsistent} with a principled (e.g., probabilistic) approximation that leverages the essential physical assumptions about occlusions sketched above. 

A fully ad-hoc strategy~\emph{(i)} directly handles track losses.
Basic strategies aim to reintroduce lost object tracks when the objects become visible to the sensor again after occlusion, e.g.,~\cite{Occlusions-dGLMB:2022,Occlusions-GLMB:2020,Baisa:GMPHD-Occlu:2021,Specker:OccluMultiReID:2021}.
One can also compute a variable reflecting the current status of occlusions (e.g., visibility) and use it to prevent the method from losing the track in the first place.
In the domain of computer vision~(CV), this can be achieved by utilizing \mbox{re-identification} (re-ID) visual features, e.g.,~\cite{Baisa:GMPHD-Occlu:2021,Specker:OccluMultiReID:2021,BoT-SORT:2022,FastTracker:2025}.
Other approaches from the CV domain include ByteTrack~\cite{ByteTrack:2021}, which uses \emph{all} bounding box outputs (i.e., detections) of a visual detection network, including the detections with low \emph{scores}.
Methods~\cite{SparseTrack:2023} or~\cite{A-p3DKF:2023} employ so-called pseudo depth and use a custom detection-to-track matching.
To handle track loss using the motion capture sensor, \cite{Occlu-LMIPDA-MarkovChainTwo:2023} employed the linear MOT approach~\cite{Musicki:LinearMultitarget:2005}, in which measurements appearing nearby other objects are considered as clutter.

Computation of variables reflecting occlusions may involve a model, which leads to strategies~\emph{(ii)}.
A number of methods employ a customary likelihood function~\cite{Wu:OccluParts:2006,StereoCam3DTracking:2009,MonocularMTT3dOcclu:2013,Yingdong:OccluSegment:2010,Ercan:OccluLikelihood:2007,Occlusions-occluState:2014,Occlusions-MHT:2014,Wyffels:NegativeOcclu:2015,Occlusions-particle:2015}.
In CV, the works \cite{Wu:OccluParts:2006,MonocularMTT3dOcclu:2013} compute \emph{visibility}, while~\cite{Yingdong:OccluSegment:2010} retain \emph{occlusion rate} variables for each object. 
Similarly, \cite{Ercan:OccluLikelihood:2007,Occlusions-occluState:2014} use an \emph{occlusion relationship} variable, which approach shares similarity with~\cite{StereoCam3DTracking:2009}.
The works~\cite{Occlusions-MHT:2014,Wyffels:NegativeOcclu:2015} consider missing measurements due to occlusions as negative information. 
The work~\cite{Occlusions-particle:2015} employs particle filters that interact via a likelihood function accounting for occlusions.
Instead of defining a custom likelihood, one can craft an \emph{association cost matrix} directly~\cite{Monocular3dHead:2019,ByteTrack:2021,BoT-SORT:2022}, based on occlusion\discretionary{-}{-}{-}related variables.
Occluded regions are computed in~\cite{Occlusions-PHD:2014} for radar tracking, where objects whose mean vector falls into the occlusion region of the other objects are programatically retained in the algorithm.
A similar approach is utilized in~\cite{Chen:LidarOccluMap:2018} for LiDAR tracking.

The ad-hoc strategies \emph{(i)-(ii)} may be reasonable and lead to superior performance.
However, the heuristic data processing involved in these methods may lead to doubts about their application in safety-critical areas.
To alleviate this, a fully model-based approach \emph{(iii)} can be used.
The method~\cite{Occlusions-visual:2011} proposed minimizing a custom physical \emph{energy} that included an occlusion model, which shares similarities with~\cite{GMPHDOGM17:2019}.
A more common approach, however, is to formulate the entire physical model in probabilistic terms and use the Bayes rule to yield the corresponding optimal solution~\cite{Mahler-Book:2007,Mahler-Book:2014,Streit-Book:2021}.
Note that despite adopting the Bayesian formulation, the methods~\cite{Occlusions-GLMB:2020,Baisa:GMPHD-Occlu:2021,Occlusions-dGLMB:2022} fall into~\emph{(i)} due to their occlusion-handling strategy.

To account for occlusions in the Bayesian formulation consistently, one can either \emph{(iii.a)} adopt the merged measurement framework~\cite{Kim-LRFS-Occlu-Merged:2019,Merged-RFS:2015,clark2012faa}, or \emph{(iii.b)} model probability of detection (\probabilityOfDetection{}) as a \emph{function} reflecting occlusions.
The latter is more common in the literature.
While \emph{(iii.a)} seems reasonable for the track\discretionary{-}{-}{-}before\discretionary{-}{-}{-}detect paradigm used in~\cite{Kim-LRFS-Occlu-Merged:2019}, it seems unsuitable for tracking\discretionary{-}{-}{-}by\discretionary{-}{-}{-}detection focused on by this work, where a detection can usually be assigned to one object only.
Outside the Bayesian formulation, a related strategy based on object groups is taken in~\cite{Yang:OccluGroup:2005,Senior:OccluGroup:2006}, where objects participating in an occlusion event were merged together.

Drastic approximations involved in existing methods of~\emph{(iii.b)} make them fall into~\emph{(ii)}.
Despite the \emph{design} of the \probabilityOfDetection{} function might be sound, its integration into the method follows ad-hoc workarounds.
The work~\cite{Occlusions-PHD-EOT:2012} assigns each object a \probabilityOfDetection{} assuming it is a constant function of its mean vector, while~\cite{Occlusion-PMBM-MeasWise:2018,Occlusions-PHD:2015,Strand:Occlu:2023,Occlusions-GLMB-Vo:2022,OccluGLMB:multiView:2024,Linh:VisualOccluLMB:2024} assume it depends on an occlusion\discretionary{-}{-}{-}related variable. 
In particular,~\cite{Occlusions-PHD:2015} extends~\cite{Occlusions-PHD:2014} by using an \emph{occlusion likelihood} that scales the \probabilityOfDetection{}.
The studies~\cite{Occlusions-GLMB-Vo:2022,OccluGLMB:multiView:2024,Linh:VisualOccluLMB:2024} model the \probabilityOfDetection{} for each object as being directly influenced by the other objects. 
This allowed~\cite{Occlusions-GLMB-Vo:2022} to conclude that the posterior is no longer conjugate with the prior used therein, and thus intractable.
As a simple workaround, \cite{Occlusions-GLMB-Vo:2022,OccluGLMB:multiView:2024,Linh:VisualOccluLMB:2024} estimated the states of other objects and substituted them into the \probabilityOfDetection{}. 
This is arguably a sound approximation, provided the objects have little spatial and existence uncertainties.
The uncertainties, however, usually vary in time and space. 

A more sound approximation given in~\cite{Dai-Occlusion:2020} took the uncertainties into account by a weighted sum over data association hypotheses.
After analyzing~\cite{Dai-Occlusion:2020}, we conclude that the notation and interpretation of the procedure are questionable.
Similarly, the definition of \emph{occlusion probability} used to scale the \probabilityOfDetection{} in~\cite{Occlusion-PMBM-MeasWise:2018} is mathematically vague and thus questionable.
Finally, the works~\cite{Occlusions-CPHD:2012,Occlusions-CPHD:2013} designed the \probabilityOfDetection{} to account for both occlusions and the existence of all the objects \emph{explicitly}.
To use it efficiently for each \emph{heuristically} indexed object, an \emph{expected} \probabilityOfDetection{} (\expectedProbabilityOfDetection) was computed over its distribution. 
Note that ad-hoc indexing is a subtle issue in a number of the above methods~\cite{Occlusions-PHD:2014,Occlusions-PHD:2015,Occlusion-PMBM-MeasWise:2018,Strand:Occlu:2023,Occlusions-CPHD:2012,Occlusions-CPHD:2013}.

To the best of the authors' knowledge, none of the existing methods handle occlusions in a fully model-based and thus consistent manner \emph{(iii)}, while also resulting in a tractable end\discretionary{-}{-}{-}to\discretionary{-}{-}{-}end algorithm at the same time. 
This paper addresses this issue by developing a domain-independent variational\discretionary{-}{-}{-}Bayesian solution, that yields a tractable occlusion-handling strategy.
The main contributions of this paper are:
\begin{itemize}
    \item
    Two approximations to the multi-object posterior are derived, given a measurement model whose \probabilityOfDetection{} depends on all objects.
    The corresponding occlusion-handling strategies involve computing the \expectedProbabilityOfDetection{}
    for each (\emph{1}) state or (\emph{2}) mark, and the expectation is taken over the \emph{reduced Palm density} (RPD)\cite{Vere-Jones:2008}.
    \item 
    The RPD computation is analytically revealed for several multi-object densities used in MOT problems. 
    \item 
    Tractable \expectedProbabilityOfDetection{} computation is proposed for the prior being a multi-Bernoulli mixture (MBM) with marks.
    \item
    The MBM filter with the proposed occlusion-handling strategy is applied to visual tracking and shown to outperform comparable existing methods. 
\end{itemize}

The paper is organized as follows.
Section~\ref{sec:background-RFS} introduces background on Bayesian MOT needed for the proposed solution.
Section~\ref{sec:proposed-general} introduces Palm conditioning, used when a particular object is known to be present, and auxiliary variables used for convenient measurement density representation.
Then the proposed method is outlined.
Section~\ref{sec:special-case} presents a more efficient approximation and a filter implementation.
The proposed method is tested in Section~\ref{sec:Results}, and the paper concludes in Section~\ref{sec:Conclusion}.

\section{Background of Model-Based MOT}\label{sec:background-RFS}
This section briefly reviews \emph{random finite set} (RFS) approach to MOT with Bayesian treatment of \emph{finite set statistics} (FISST)~\cite{Mahler-Book:2007,Mahler-Book:2014}, with special regard to the so-called \emph{standard point-object} (SPO) model. 
Furthermore, the occlusion-handling strategy of~\cite{Occlusions-GLMB-Vo:2022,OccluGLMB:multiView:2024,Linh:VisualOccluLMB:2024} is reviewed.

\subsection{RFS Approach to MOT}
Throughout the paper, the symbols $\X$ and $\Z$ denote the state and measurement spaces, respectively, and are both assumed to be \emph{complete separable metric} and measure spaces with some reference measures (such as the Lebesgue measure in the case of $\X {=} \mathbb{R}^d$). 
For clarity, unnecessary mathematical details are omitted if possible.

Intuitively, an RFS $\Xi$ on $\X$ is a random subset of $\X$ with a locally finite number of elements (analogically for $\Z$).
That is, both the cardinality $|\Xi|$ and the elements $\mathbf{x} {\in} \Xi$ are random.
The elements are regarded as \emph{points}.

Throughout the paper, the symbol $\StateSet_k {=} \{ \mathbf{x}_k^1,\dots,\mathbf{x}_k^{n_k} \}$ denotes the set of object states at time step $k{=}0,\dots,K$, with $K {\in} \mathbb{N}$ being the final time step.
That is, $X_k$ is a realization of $\Xi_k$, $k{=}0,\dots,K$.
Whenever possible, random entities and their realizations are denoted by the same symbol for convenience.
Similarly, measurements received at each time step are modeled as RFSs $Z_k {=} \{ \mathbf{z}_k^1, \dots, \mathbf{z}_k^{m_k}\}$. 

For $k{=}0,\dots,K$, the available measurement sets $Z^k {\triangleq} (Z_0,\dots,Z_k)$ are used to yield the posterior density $p(X_k|Z^k)$ of $X_k$ via the Bayes filter~\cite{Mahler-Book:2007}
\begin{subequations}\label{eqs:MTT-Bayes-filter}
\begin{align}
    p \big(X_k | Z^k \big) &\propto p(Z_k | X_k) \cdot p(X_k | Z^{k-1}), \label{eq:MTT-Bayes-filter-Bayes}\\
    \! p(X_{k} | Z^{k-1}) &= \textstyle\int p(X_{k}|X_{k-1}) p(X_{k-1}|Z^{k-1}) \delta X_{k-1}  , \!\!
\end{align}
\end{subequations}
where $p(X_0|Z^{-1}) {\triangleq} p(X_0)$ is the initial density for $k{=}0$, $p (X_{k+1} | Z^{k})$ is the predictive density and the densities $p(Z_k | X_k)$ and $p(X_{k+1}|X_{k})$ encapsulate the measurement and motion models (such as the SPO model), respectively.
The symbol $\propto$ denotes equality up to a normalizing constant.
The subscript of density functions will be used to provide additional information, such as the distribution name and/or the random variable symbol, if its omission could lead to confusion. 
The integral of a function $f(X)$ with finite-set inputs is taken to be the \emph{set integral}~\cite{Mahler-Book:2007}
\begin{align}    
    \! \textstyle \int \! f(X) \delta X \triangleq \sum_{n=0}^{+\infty} \tfrac{1}{n!} \! \int \! f (\{ \mathbf{x}^1,\dots,\mathbf{x}^n \}) \mathrm{d}\mathbf{x}^1\cdots\mathrm{d}\mathbf{x}^n \! , \! \label{eq:FISST-set-integral}
\end{align}
and the densities are such that $\int p(X)\delta X = 1$.
Several key RFS densities are listed below.

\subsubsection{Bernoulli RFS}
A Bernoulli RFS is either empty or a singleton.
The corresponding density function is
\begin{align}
    p_{  \mathrm{Ber}}(\StateSet) = \begin{cases}
        1-r & \text{if } \StateSet = \emptyset, \\[-0.1cm]
        r \cdot p_{\mathrm{sp}}(\mathbf{x}) & \text{if } \StateSet = \{\mathbf{x}\}, \\[-0.1cm]
        0 & \text{otherwise},
    \end{cases}
    \label{eq:Bernoulli-RFS-density}
\end{align}
where $r\in[0,1]$ is the existence probability and $p_{\mathrm{sp}}(\mathbf{x})$ is the spatial density of $\mathbf{x}$ conditioned on its existence.

\subsubsection{Multi-Bernoulli RFS}
A multi-Bernoulli (MB) RFS is a union of multiple independent Bernoulli RFSs.
Denoting the density of the $\ell$-th Bernoulli RFS with $p_{\mathrm{Ber}}^\ell (X)$, $\ell{=}1,\dots,N$, the resulting MB RFS density is given by
\begin{align}
    p_{ \mathrm{MB} }(\StateSet) = \textstyle \sum_{ \uplus_{\ell=1}^{N} Y^\ell = \StateSet} \, \prod_{\ell = 1}^N p_{  \mathrm{Ber} }^\ell ( Y^\ell ) \, ,
    \label{eq:MB-density-function}
\end{align}
where the symbol $\uplus$ means that the sum is taken over all mutually \emph{disjoint} sets $Y^1,\dots,Y^N$ whose union is $\StateSet$.

\subsubsection{Poisson RFS}
A Poisson RFS describes an independent and identically distributed set of points whose number is Poisson distributed.
The corresponding density function is
\begin{align}
    p_{ \mathrm{Pois} } (\StateSet) = \textstyle \e^{-\rho } \prod_{\mathbf{x} \in \StateSet} \, \rho \cdot p_{\mathrm{sp}}(\mathbf{x}) \, ,
    \label{eq:Poisson-density-function}
\end{align}
where $\rho \geq 0$ is the expected number of points.

\subsubsection{Poisson Multi-Bernoulli RFS}
The union of independent Poisson-distributed RFS $X^0$ and MB-distributed RFS $X^1$ is the Poisson multi-Bernoulli (PMB) RFS.
The corresponding density function is
\begin{align}
    p_{\mathrm{PMB}}(X) = \textstyle \sum_{ Y^0 \uplus Y^1 = \StateSet} \, p_{\mathrm{Pois}} \big(Y^0\big) \cdot p_{  \mathrm{MB} } \big( Y^1 \big) \, , \label{eq:PMB-density-function}
\end{align}
where the summation is analogical to that in~\eqref{eq:MB-density-function}.

\subsubsection{Marked RFS}
As a special RFS case, an RFS $X {\subset} \X$ with $\X {=} \X^{\projFcn} {\times} \spaceOfMarks$ being a joint space consisting of a complete separable metric space $\X^{\projFcn}$ and a set of \emph{marks} $\spaceOfMarks$ is called a \emph{marked RFS}, whenever the points without marks (i.e., the corresponding marginal random set on $\X^\projFcn$) again forms an RFS, c.f.~\cite[Sec.~1.8]{StochasticGeometry-lecturesBook:2004}.
In this paper, the \emph{space of marks} $\spaceOfMarks$ is assumed to be discrete.
Analogically for $\Z$.

Points with marks will be denoted as $\mathbf{x} {=} (x,m) {\in} \X$.
To extract marks from a set $X$ of marked points, define
\begin{align}
    \mathcal{L}( X ) &\triangleq \{ m {\in} \spaceOfMarks : \, \exists x {\in} \X^\projFcn \text{ s.t. } (x,m) {\in} X \} \, .
\end{align}
Note that marks are not necessarily unique, i.e., $|\mathcal{L}( X )|$ might not be equal to $|X|$.
For a set $X$, let 
\begin{align}
    X^m \triangleq \big\{ \mathbf{x} {\in} X : \, \mathcal{L}( \{ \mathbf{x} \} ) {=} \{m\}  \big\} \, ,
    \label{eq:points-with-same-marks}
\end{align}
denote the subset of points of $X$ with the same mark $m {\in} \spaceOfMarks$.
Integral of a function $f(\mathbf{x})$ on $\X {=} \X^{\projFcn} {\times} \spaceOfMarks$ is 
\begin{align}
    \textstyle \int f(\mathbf{x}) \d \mathbf{x} = \sum_{m\in\spaceOfMarks} \int f(x,m) \d x \, ,
\end{align}
and the set integral~\eqref{eq:FISST-set-integral} of a function $f(X)$ with finite-set inputs thus takes the form
\begin{align}
    & \textstyle  \int f(X) \delta X = \sum_{n=1}^{+\infty} \frac{1}{n!} \sum_{m^1,\dots,m^n\in\spaceOfMarks} \notag \\
    & \hspace{1.1cm} \textstyle  \int f( \{(x^1,m^1), \dots, (x^n,m^n)\} ) \d x^1 \cdots \d x^n \, . 
\end{align}
Denoting the density of a marked RFS with $p_{ \mathrm{marked} }(\,\cdot\,)$, the corresponding set of points without marks has the marginal density for any $\{ x^1, \dots, x^n \} {\subset} \X^\projFcn$ given by
\begin{align}
    \label{eq:demarking-a-density}
    \! & p_{ \mathrm{unmarked} } ( \{ x^1, \dots, x^n \} ) = \notag \\[-0.1cm]
    & \hspace{0.13cm} \textstyle \sum_{ m^1, \dots, m^n \in \spaceOfMarks } p_{ \mathrm{marked} } ( \{ (x^1,m^1), \dots, (x^n,m^n) \} ) . \!
\end{align}

A significant portion of the MOT literature is concerned with the case when points must always have distinct marks.
In such a case, the marks are referred to as \emph{labels}~\cite{VoVo-GLMB:2011,VoVo-dGLMB:2013,VoVo-dGLMB:2014} and are used to distinguish among different objects to form trajectories.
Labels typically apply to a set of states only, while in this paper, both the sets of states and measurements will be considered as having marks.

\subsection{MOT Without the Presence of Occlusions}\label{sec:MOT-without-occlusions}
A widely adopted model for tracking objects without any interactions (including occlusions) is given by the SPO model.
It should be emphasized that the terminology \emph{point}-object refers to the assumption that each point (i.e., object) $\mathbf{x}_k {\in} X_k$ is \emph{resolvable} as an actual \emph{point} for the sensor.
That is, each object is either detected or not, whereas there is at most one detection per object. 

In the SPO measurement model,
each object $\mathbf{x}_k {\in} \StateSet_{k}$ is either detected with the state-dependent \probabilityOfDetection{} $P_\ObjectGenRFS(\mathbf{x}_k)$ and independently generates a measurement according to the single-object likelihood function $\likelihood(\mathbf{z}_k|\mathbf{x}_k)$, or it is undetected with the probability $1 {-} P_\ObjectGenRFS(\mathbf{x}_k)$.
Moreover, independent clutter detections described with the RFS~$\ClutterRFS_k$ may be produced by a sensor.
The set of all measurements at time step $k$ is thus~\cite[pp.~411]{Mahler-Book:2007},
\begin{align}\label{eq:MTT_MeasurementRFS_Standard-point-object}
	Z_k =
            \big( \, \textstyle \bigcup_{\mathbf{x}_k\in\StateSet_k} \ObjectGenRFS(\mathbf{x}_k) \, \big) 
            \cup
            \ClutterRFS_k \, ,
\end{align}
where $\ObjectGenRFS(\mathbf{x}_{k})$ is a conditional Bernoulli RFS with density
\begin{align}\label{eq:standard-point-object_measurement-object-gen_Bernoulli}
	p(D|\mathbf{x}_{k}) = \begin{cases}
		1-P_\ObjectGenRFS(\mathbf{x}_{k}) & \text{if } D = \emptyset \, \\[-0.1cm]
		P_\ObjectGenRFS(\mathbf{x}_{k})\likelihood(\mathbf{z}_k | \mathbf{x}_{k} ) & \text{if } D = \{\mathbf{z}_k\} \,, \\[-0.1cm]
		0 & \text{otherwise}.
	\end{cases}
\end{align}
The clutter RFS $\ClutterRFS_k$ is modeled as Poisson.
The density $p(Z_k|X_k)$ resulting from~\eqref{eq:MTT_MeasurementRFS_Standard-point-object} is thus a PMB~\eqref{eq:PMB-density-function}~\cite[Sec.~12]{Mahler-Book:2007}.
The SPO motion model is omitted in this paper for brevity, and can be found, e.g., in~\cite[pp.~467]{Mahler-Book:2007}.

Assuming the birth and the initial density are both MB (or mixtures thereof), the corresponding model-based filter is the multi-Bernoulli mixture (MBM) filter~\cite{Mahler-integralTransform:2016,GM-MBM:2019}.
If object states for the MBM filter are uniquely marked, the labeled MBM (LMBM) filter~\cite{Mahler-integralTransform:2016} results without the need to alter the filtering recursion.
The so-called generalized labeled multi-Bernoulli (GLMB)~\cite{VoVo-dGLMB:2014}, also referred to as $\text{MBM}_{01}$ in~\cite{PMBM:2018}, can be understood as a special case of the LMBM filter, where the existence probability of each Bernoulli is either 0 or 1, resulting in an exponential increase of global hypotheses in the prediction step.
Approximations can be employed to yield, e.g., the LMB~\cite{LMB:2014} and other filters~\cite{Williams-MarginalPMBM:2015}.
Assuming the birth and the initial distribution are Poisson is possible as well, see, e.g.,~\cite[Section~16.3]{Mahler-Book:2007}, \cite{Williams-MarginalPMBM:2015,PMBM:2018}.

Particular filtering recursions are omitted and can be found in the literature. 
Note that \emph{Gaussian mixture} (GM) implementations often assume constant \probabilityOfDetection{} $P_\ObjectGenRFS(\mathbf{x}_k) {=} P_\ObjectGenRFS$.

\subsection{MOT in the Presence of Occlusions}\label{sec:accounting-for-occlusions}
In this section, the SPO model is generalized to account for occlusions following~\cite{Occlusions-GLMB-Vo:2022,OccluGLMB:multiView:2024,Linh:VisualOccluLMB:2024}.
In particular, the structure of the SPO measurement model~\eqref{eq:MTT_MeasurementRFS_Standard-point-object}-\eqref{eq:standard-point-object_measurement-object-gen_Bernoulli} is adjusted to allow the possibility that detections are less likely to occur for occluded objects.
The corresponding model is referred to as SPO with \emph{dependency} (SPO-D).

Both the \probabilityOfDetection{} and the single-object likelihood function for each $\mathbf{x}_k {\in} X_k$ may also depend on the rest of the objects $X_k {\setminus} \{\mathbf{x}_k\}$.
Since we are modeling \emph{point objects}, it is reasonable to assume that measurements, if received, are not influenced by the presence of other objects.
Therefore, the dependency of the likelihood function $\likelihood(\mathbf{z}_k | \mathbf{x}_{k} )$ on the other objects $X_k {\setminus} \{\mathbf{x}_k\}$ is dropped.
That is, Eq.~\eqref{eq:MTT_MeasurementRFS_Standard-point-object} is generalized for the SPO-D model to
\begin{align}\label{eq:MTT_MeasurementRFS_Standard-point-object-with-dependency}
	Z_k =
            \big( \, \textstyle \bigcup_{\mathbf{x}_k\in\StateSet_k} \ObjectGenRFS(\mathbf{x}_k,\, X_k {\setminus} \{ \mathbf{x}_k \}) \, \big) 
            \cup
            \ClutterRFS_k \, ,
\end{align}
where $\ObjectGenRFS(\mathbf{x}_{k}, \, X_k {\setminus} \{ \mathbf{x}_k \})$ is a Bernoulli RFS with 
\begin{align}\label{eq:standard-point-object_measurement-object-gen_Bernoulli-with-dependency}
	&p(D \, | \, \mathbf{x}_{k}, X_k {\setminus} \{ \mathbf{x}_k \}) \\[-0.1cm]
    &\hspace{0.7cm}= \begin{cases}
		1-P_\ObjectGenRFS(\mathbf{x}_{k}, X_k {\setminus} \{ \mathbf{x}_k \}) & \text{if } D = \emptyset\, , \\[-0.1cm]
		P_\ObjectGenRFS(\mathbf{x}_{k}, X_k {\setminus} \{ \mathbf{x}_k \}) \cdot \likelihood(\mathbf{z}_k | \mathbf{x}_{k} ) & \text{if } D = \{\mathbf{z}_k\} \, , \\[-0.1cm]
		0 & \text{otherwise},
	\end{cases}
    \notag
\end{align}
where $P_\ObjectGenRFS(\mathbf{x}_{k}, X_k {\setminus} \{ \mathbf{x}_k \})$ is the \probabilityOfDetection{} that depends on the given object $\mathbf{x}_k$ and the set of \emph{other} objects $X_k {\setminus} \{\mathbf{x}_k\}$.
Examples of $P_\ObjectGenRFS(\mathbf{x}_{k}, X_k {\setminus} \{ \mathbf{x}_k \})$ expressions can be found in~\cite{Occlusions-GLMB-Vo:2022,OccluGLMB:multiView:2024,Linh:VisualOccluLMB:2024} and will be discussed in detail later.
Clutter $\ClutterRFS_k$ is still modeled as Poisson.
It is easy to see that the density $p(Z_k|X_k)$ resulting from~\eqref{eq:MTT_MeasurementRFS_Standard-point-object-with-dependency} is still a PMB~\eqref{eq:PMB-density-function}, since $X_k$ is given.
The SPO-D motion model is taken to be the SPO motion model.

Unless the prior is Bernoulli, the posterior corresponding to the SPO-D model contains \emph{interactions} among objects induced by the dependency of $D(\mathbf{x}_k, X_k {\setminus} \{\mathbf{x}_k\})$ on $X_k {\setminus} \{\mathbf{x}_k\}$.
This was observed by~\cite{Occlusions-GLMB-Vo:2022} for the GLMB prior.
To achieve tractability of the posterior, approximations must be developed.
The works~\cite{Occlusions-GLMB-Vo:2022,OccluGLMB:multiView:2024,Linh:VisualOccluLMB:2024} used an ad-hoc approximation within the labeled framework, which could be stated as follows:

\def\solOne{E} 
\def\solTwo{S} 
\def\solThree{O} 
\def\solVo{\mbox{\solOne\solTwo\solThree}}

\def\ourOne{P} 
\def\ourTwo{R} 
\def\ourThree{O} 
\def\ourName{\ourOne\ourTwo\ourThree}

\begin{itemize}
    \item[\solOne:] estimate the set of all objects $\hat{X}_k$ based on the prior,
    \item[\solTwo:] substitute $\hat{X}_k$ into the SPO-D model, i.e., define
    \begin{align}
        P_\ObjectGenRFS( m ) &\coloneqq P_\ObjectGenRFS\big( ( \hat{x}_k, m) , \, \hat{X}_k {\setminus} \{ ( \hat{x}_k, m) \}\big) \,, 
        \label{eq:approx-Vo2022}
    \end{align}
    where $\hat{x}_k$ is the estimate corresponding to label $m$,
    \item[\solThree:] obey the SPO model parameterized with~\eqref{eq:approx-Vo2022} to achieve tractable filtering recursion.
\end{itemize}
This occlusion-handling strategy will be referred to as \solVo{} strategy for short.
Since the estimation step \solOne{} is based on existence probabilities,
the \solVo{} strategy considers (\emph{I}) no spatial uncertainties and (\emph{II}) only a limited amount of existence uncertainties corresponding to the objects.
Note that the computational complexity of \solVo{} depends on the chosen estimator and on the  \probabilityOfDetection{} function implementation.

Other strategies that take the uncertainties into account are either insufficiently described~\cite{Occlusion-PMBM-MeasWise:2018,Dai-Occlusion:2020} or assume that the \probabilityOfDetection{} depends on the uncertainties explicitly~\cite{Occlusions-CPHD:2012,Occlusions-CPHD:2013}, which fails to follow physical modeling.
A strategy is proposed in the following section that considers uncertainties in a mathematically sound manner.

\section{Proposed Approximation}\label{sec:proposed-general}
The proposed occlusion-handling strategy modifies the steps \solOne{}\emph{} and \solTwo{}\emph{}, while it keeps the appealing step \solThree{}\emph{} in use.
It serves as an approximation of the posterior, which will be detailed later.
Specifically, Eq.~\eqref{eq:approx-Vo2022} is replaced by calculating a conditional \emph{expectation} of $P_\ObjectGenRFS(\mathbf{x}_{k}, X_k {\setminus} \{ \mathbf{x}_k \})$, i.e., \expectedProbabilityOfDetection{}.
Note that~\cite{Occlusions-CPHD:2012} computed an \expectedProbabilityOfDetection{} differently, and that \expectedProbabilityOfDetection{} can be found in other contexts~\cite{SongMusicki:PdStateDep:2012,Jones:NonMyopicSensorManagement:2024}.

To theoretically develop the strategy outlined above, we first introduce \emph{Palm conditioning} (Section~\ref{sec:palm}) and \emph{auxiliary variables} 
(Section~\ref{sec:auxiliary-variables}), which are essential for establishing the approximation in Section~\ref{sec:EPD}.\\[-0.7cm]

\subsection{Palm Conditioning}\label{sec:palm}
The \emph{Palm distribution} is the conditional probability law of the RFS $\Xi$, given that a particular point $\mathbf{x} {\in} \X$ is known to be an element of $\Xi$, i.e., given the event $\mathbf{x} {\in} \Xi$.
The \emph{reduced} Palm distribution is then the probability law of the \emph{other} points, i.e., of the subset $\Theta {=} \Xi {\setminus} \{\mathbf{x}\}$, given $\mathbf{x} {\in} \Xi$.
Its corresponding density function, called the \emph{reduced Palm density} (RPD) is needed in this paper.
In the FISST notation, the RPD evaluated at $O \coloneqq X {\setminus} \{\mathbf{x}\}$ can be computed simply as (see~\cite[p.~511]{Vere-Jones:2008})
\begin{align}
     p_{\Theta | \mathbf{x}}( O | \mathbf{x}) \overset{\text{abbr.}}{=} p_{\Xi \setminus \{\mathbf{x}\} | \mathbf{x}\in\Xi } ( O | \mathbf{x} ) = \frac{ p_{\Xi} (O \cup \{\mathbf{x}\}) }{ D_{\Xi}(\mathbf{x}) } \label{eq:RPD-RFS-definition}
    \,, 
\end{align}
where $D_{\Xi}(\mathbf{x})$ is the PHD corresponding to $\Xi$ as
\begin{align}
    D_{\Xi}(\mathbf{x}) = \textstyle \int p_{\Xi}(X \cup \{\mathbf{x}\}) \delta X \label{eq:PHD-definition} \, .
\end{align}
Palm conditioning springs from the random counting measure treatment of RFSs, which is an alternative to FISST.
The rigorous construction of~\eqref{eq:RPD-RFS-definition} within FISST is detailed in Appendix~\ref{app:Palm-definition}.
Several RPD examples follow.\\[-0.5cm]

\subsubsection{RPD Corresponding to a Bernoulli RFS}
It follows trivially that the RPD for a Bernoulli RFS is equal to one if $O {=} \emptyset$ and zero otherwise.
It also follows from~\eqref{eq:RPD-RFS-definition} using the Bernoulli PHD $D_{\mathrm{Ber}}(\mathbf{x}) {=} r {\cdot} p_{\mathrm{sp}}(\mathbf{x})$.\\[-0.5cm]

\subsubsection{RPD Corresponding to a Poisson RFS}
According to the \emph{Slivnyak-Mecke} theorem~\cite[p.~281]{Vere-Jones:2008}, Palm conditioning within a Poisson process introduces no knowledge of the distribution of the other points. 
Indeed, the RPD~\eqref{eq:RPD-RFS-definition} for the Poisson RFS~\eqref{eq:Poisson-density-function} is 
\begin{align}
    \! p_{ \mathrm{Pois:}\Theta | \mathbf{x} } ( O | \mathbf{x} ) {=} \frac{ \e^{-\rho} \prod_{\mathbf{o} \in O \cup \{\mathbf{x}\} } \rho {\cdot} p_{\mathrm{sp}}(\mathbf{o}) }{ \rho {\cdot} p_{\mathrm{sp}}(\mathbf{x}) } {=} p_{\mathrm{Pois}} (O) . \!
\end{align}

\subsubsection{RPD Corresponding to a Poisson Multi-Bernoulli RFS}
Consider an RFS $\Xi$ with a PMB density~\eqref{eq:PMB-density-function}.
A point $\mathbf{x} {\in} \Xi$ might have come from either the Poisson component or one of the Bernoulli components.
Conditioning on ``$\mathbf{x} {\in} \Xi$'' must thus explore either possibility, leading to a mixture of PMB densities.
First, the numerator of~\eqref{eq:RPD-RFS-definition} is
\begin{align}
    \!\! p_{\mathrm{PMB}} &(O {\cup} \{\mathbf{x}\}) \!=\! \textstyle \sum_{ O^0 \uplus O^1 = O \cup \{ \mathbf{x} \} } p_{\mathrm{Pois}}(O^0) {\cdot} p_{\mathrm{MB}} (O^1) \notag \\
    &= \!\!\! \textstyle \sum_{ Y^0 \uplus Y^1 = O } p_{\mathrm{Pois}}(Y^0) \Big( \rho {\cdot} p_{\mathrm{sp}}(\mathbf{x}) {\cdot} p_{\mathrm{MB}} (Y^1) \notag\\[-0.12cm]
    & \hspace{2.1cm} + \textstyle \sum_{\ell=1}^{N} r^{\ell} {\cdot} p_{\mathrm{sp}}^\ell(\mathbf{x}) {\cdot} p_{\mathrm{MB} \setminus \ell} (Y^1 ) \Big) , \!
\label{eq:RPD-for-PMB-derivation}
\end{align}
where $r^\ell$ is the existence probability and $p_{\mathrm{sp}}^\ell(\mathbf{x})$ is the spatial probability density corresponding to the $\ell$\discretionary{--}{-}{-}th Bernoulli component, respectively; $\ell{=}1,\dots,N$,
The symbol $\mathrm{MB}{\setminus}\ell$ is a shorthand notation for the MB resulting from omitting the $\ell$-th component, 
\begin{align}
    p_{\mathrm{MB} \setminus \ell} (X) = 
    \textstyle \sum_{ \uplus_{ \substack{ m=1 \\ m \neq \ell } }^N Y^m = Y } 
    \prod_{ \substack{ \textcolor{white}{.}\\[0.0cm] m=1 \\ m\neq\ell } }^N p^m_{\mathrm{Ber}}(Y^m) \,. \label{eq:MB-without-ell}
\end{align}
Dividing~\eqref{eq:RPD-for-PMB-derivation} with the PHD
\begin{align}
    D_{ \mathrm{PMB} } (\mathbf{x}) &= \rho {\cdot} p_{\mathrm{sp}}(\mathbf{x}) + \textstyle \sum_{\ell=1}^{N} r^{\ell} {\cdot} p_{\mathrm{sp}}^\ell(\mathbf{x})
\end{align}
(c.f.~\eqref{eq:RPD-RFS-definition}) yields the desired RPD as 
\begin{align}
    p_{ \mathrm{PMB:} \Theta | \mathbf{x} } (O|\mathbf{x}) &= \textstyle
    \sum_{ O^0 \uplus O^1 = O } p_{\mathrm{Pois}}(O^0) \big( \, w^0 {\cdot} p_{\mathrm{MB} } (O^0) \notag\\[-0.12cm]
    & \hspace{1cm} + \textstyle \sum_{\ell = 1}^N w^\ell {\cdot} p_{\mathrm{MB}\setminus \{\ell\} } (O^1) \, \big) \, ,
    \label{eq:RPD-for-standard-PMB-density}
\end{align}
which is a PMB mixture, where the weights are 
\begin{subequations}\label{eq:RPD-for-standard-PMB-weights}
\begin{align}
    w^0 &= \tfrac{ \rho {\cdot} p_{\mathrm{sp}}(\mathbf{x}) }{ \rho {\cdot} p_{\mathrm{sp}}(\mathbf{x}) + \sum_{\ell=1}^{N} r^{\ell} {\cdot} p_{\mathrm{sp}}^\ell(\mathbf{x}) } \, , \\[-0.05cm]
    w^\ell &= \tfrac{ r^{\ell} {\cdot} p_{\mathrm{sp}}^\ell(\mathbf{x}) }{ \rho {\cdot} p_{\mathrm{sp}}(\mathbf{x}) + \sum_{\ell=1}^{N} r^{\ell} {\cdot} p_{\mathrm{sp}}^\ell(\mathbf{x}) } \, , \hspace{0.8cm} \ell = 1,\dots,N .
\end{align}
\end{subequations}
Note that the RPD corresponding to MB RFS is a special case of~\eqref{eq:RPD-for-standard-PMB-density}-\eqref{eq:RPD-for-standard-PMB-weights} when setting $\rho {=} 0$.

If a point $\mathbf{x} {\in} \Xi$ contains the information from which PMB component it came from, both the PMB density and its RPD get simplified significantly.
We aim to simplify the PMB densities related to the SPO and \mbox{SPO-D} measurement likelihoods for convenience.
For this reason, a special class of marked RFSs is introduced, whose marks are made to correspond with the different density components.
Such marks effectively behave like auxiliary variables that facilitate the development of efficient approximations~\cite{GaFerSvenWillXia:TPMB:2020}.

\subsection{Auxiliary Variables}\label{sec:auxiliary-variables}
First, define a spatial density \emph{auxiliary supported on} $\mu {\in} \spaceOfMarks$ to be a spatial density on the joint space $\X {=} \X^{\projFcn} {\times} \spaceOfMarks$, for which 
\begin{align}
    p_{ \mathrm{sp} } ( x,m ) = p_{ \mathrm{sp} } ( x | m ) \cdot \delta_{\mu}[m] \, , \hspace{0.5cm} (x,m)\in\X \, ,
\end{align}
where $p_{ \mathrm{sp} } ( x | m )$ is a spatial density on $\X^\projFcn$ conditioned on $m {\in} \spaceOfMarks$, and $\delta_{\mu}[m]$ is the Kronecker delta.
Moreover, an \emph{auxiliary Bernoulli (resp. Poisson) RFS supported on} $\mu {\in} \spaceOfMarks$ is a Bernoulli (resp. Poisson) RFS whose spatial density is auxiliary supported on $\mu$.
The corresponding density functions are denoted with $p_{\mathrm{A}\text{-}\mathrm{Pois}}^\mu(X)$ and $p_{\mathrm{A}\text{-}\mathrm{Ber}}^\mu(X)$, for the Poisson and Bernoulli cases, respectively.

The union $X = X^0 \uplus X^1 \uplus \dots \uplus X^N$ of independent RFSs where $X^0$ is auxiliary Poisson RFS supported on $\mu^0 {\in} \spaceOfMarks$, and $X^\ell$ is auxiliary Bernoulli RFS supported on $\mu^\ell {\in} \spaceOfMarks$, $\ell {=} 1,\dots,N$, with $\mu^0,\mu^1,\dots,\mu^N$ being distinct, is the \emph{auxiliary} PMB (A-PMB) RFS.
The corresponding density function defined on the joint space is 
\begin{align}
    & p_{ \mathrm{A}\text{-}\mathrm{PMB} } (X) = \notag \\[-0.1cm]
    & \hspace{0.7cm} \textstyle
    1_{ M } ( \mathcal{L}(X) ) \cdot
    p_{\mathrm{A}\text{-}\mathrm{Pois}}^{\mu^0}\big( X^{\mu^0} \big) \prod_{ \ell = 1 }^N p_{\mathrm{A}\text{-}\mathrm{Ber}}^{\mu^\ell}\big( X^{\mu^\ell} \big) \, ,
    \label{eq:A-PMB-density-function}
\end{align}
where $M = \{ \mu^\ell \}_{\ell=1}^N$, the symbol $X^{\mu^\ell}$ is as defined in~\eqref{eq:points-with-same-marks} and $1_{A}(B)$ is the \emph{inclusion function}~\cite{VoVo-dGLMB:2013} which is equal to one if $B \subseteq A$ and zero otherwise.
The density corresponding to the set of points without marks~\eqref{eq:demarking-a-density} is the standard PMB density~\eqref{eq:PMB-density-function}, see~\cite{GaFerSvenWillXia:TPMB:2020}.

To define auxiliary measurement likelihoods, let $\Z {=} \Z^{\projFcn} {\times} (\spaceOfMarks \uplus \{0\})$ be the joint space of measurements, where $0$ will be used to determine clutter.
Consider that actual measurements received are subsets $\tilde{Z}_k$ of $\Z^\projFcn$.
Note that the SPO and SPO-D measurement models from Section~\ref{sec:MOT-without-occlusions} and~\ref{sec:accounting-for-occlusions}, respectively, are considered to be defined for $Z_k$ on $\Z$.
Moreover, it is assumed that the set of states on the joint space $\X {=} \X^{\projFcn} {\times} \spaceOfMarks$ forms a labeled RFS, i.e., elements of $X_k$ have distinct marks. 

For each given $\mathbf{x}_k {\in} \X$, specialize the definition of the single-object likelihood $\likelihood(\,\cdot\, | \mathbf{x})$ on $\Z$ to be the spatial density auxiliary supported on $\mathcal{L}(\mathbf{x}_k) {\in} \spaceOfMarks$, as
\begin{align}
    \likelihood( z,m | \mathbf{x}_k ) &= \likelihood( z | \mathbf{x}_k ) \cdot \delta_{ \mathcal{L}(\mathbf{x}_k) }[ m ]\,, \hspace{0.5cm} (z,m) \in \Z \, ,
\end{align}
where $\likelihood( z | \mathbf{x}_k )$ is the likelihood of the actual $z {\in} \Z^\projFcn$.
For the SPO measurement model, the corresponding likelihood is the A-PMB density, further called \emph{auxiliary likelihood} 
\begin{align}
    & p_{ \mathrm{A}\text{-}\mathrm{SPO}_k } ( Z_k | X_k,\, P_{\ObjectGenRFS} ) = 1_{\mathcal{L}(X_k) \uplus \{0\} } \big(\mathcal{L}(Z_k)\big) \times \notag\\[-0.1cm]
    & \hspace{1.2cm} \textstyle p_{\ClutterRFS_k} ( Z_k^{ 0 } ) \prod_{ \mathbf{x}_k \in X_k } p_{\ObjectGenRFS(\mathbf{x}_k)}( Z_k^{ \mathcal{L}(\mathbf{x}_k) } | \mathbf{x}_k,\, P_{\ObjectGenRFS} ) , 
    \label{eq:A-SPO-measurement-likelihood}
\end{align}
where $Z_k^m$~\eqref{eq:points-with-same-marks} is the subset of measurements with the mark $m$ and where $p_{\ClutterRFS_k}(\,\cdot\,)$ is a Poisson density of clutter. 
The density $p_{\ObjectGenRFS(\mathbf{x}_k)}(\,\cdot\, | \mathbf{x}_k,\, P_{\ObjectGenRFS})$ is defined in~\eqref{eq:standard-point-object_measurement-object-gen_Bernoulli}, where the dependency on $P_D(\,\cdot\,,\,\cdot\,)$ is coined explicitly. 
The SPO-D auxiliary likelihood $p_{ \mathrm{A}\text{-}\mathrm{SPO}\text{-}\mathrm{D}_k } ( Z_k | X_k )$ is a straightforward generalization of~\eqref{eq:A-SPO-measurement-likelihood} and is thus omitted.
Since~\eqref{eq:A-SPO-measurement-likelihood} is an auxiliary PMB, the density corresponding to the set of points without marks (i.e., the actual measurements) becomes the likelihood of the standard form~\eqref{eq:PMB-density-function}.
Note that the number of Bernoulli components in~\eqref{eq:A-SPO-measurement-likelihood} is $|X_k|$ and that
marks in $X_k$ must be distinct so that~\eqref{eq:A-SPO-measurement-likelihood} is A-PMB.

\subsection{SPO Approximation with Best \probabilityOfDetection{}: General Case}\label{sec:EPD}
Consider that the true MOT model obeys the \mbox{SPO-D} measurement model.
A top-down approximation of the (unnormalized) posterior is established below, using an auxiliary likelihood corresponding to a much simpler SPO model structure and an optimal parameter $P_D(\mathbf{x}_k)$.

\begin{definition}[Kullback-Leibler Divergence (KLD)]
    Given two densities $p(\,\cdot\,)$ and $q(\,\cdot\,)$ defined on a (measure) space $\mathcal{Y}$, the expression 
\begin{align}
    D_{\mathrm{KL}} (p \| q) = \textstyle \int_{\mathcal{Y}} p(y) \log \big( \tfrac{p(y)}{q(y)} \big) \, \d y \,,
\end{align}
is a statistical quantification of the discrepancy between the two densities, commonly known as the KLD. \\[-0.2cm]
\end{definition}

\newcommand{\paramPD}{\mathcal{P}}
Let $p(\,\cdot\,)$ be the true joint density on the right-hand side of~\eqref{eq:MTT-Bayes-filter-Bayes}.
That is, omitting the conditioning on $Z^{k-1}$,
\begin{align}
    p(Z_k,X_k) &\coloneqq p_{ \mathrm{A}\text{-}\mathrm{SPO}\text{-}\mathrm{D}_k } ( Z_k | X_k ) p (X_k | Z^{k-1} ) \, .
    \label{eq:true-joint}
\end{align}
Let $q(\,\cdot\,)$ be a joint density $q_\paramPD(\,\cdot\,)$ resulting from using the simpler likelihood~\eqref{eq:A-SPO-measurement-likelihood} with parameter $\paramPD : \X {\rightarrow} [0,1]$ as 
\begin{align}
    q_\paramPD(Z_k,X_k) &\coloneqq p_{ \mathrm{A}\text{-}\mathrm{SPO}_k } ( Z_k | X_k,\, \paramPD ) p (X_k | Z^{k-1} ) \, .
    \label{eq:joint-with-parameter}
\end{align}
Denote the space of joint densities of the form~\eqref{eq:joint-with-parameter} with
\begin{align}
    Q \triangleq \big\{ q_\paramPD(\,\cdot\,,\,\cdot\,) \text{ of the form~\eqref{eq:joint-with-parameter}} :\hspace{0.2cm} \paramPD : \X {\rightarrow} [0,1] \big\} .
    \label{eq:space-of-functions-with-parameter}
\end{align}
The following proposition reveals the optimal $\paramPD^*$, for which the joint densities are closest in the sense of KLD.

\begin{proposition}\label{prop:best-fitting-KLD}
    The best-fitting joint density
    \begin{align}
        q^* = \underset{ q_{\paramPD} \in Q }{ \arg\min } \ D_{\mathrm{KL}} ( p \| q_\paramPD ) \,
        \label{eq:KLD-optimality}
    \end{align}
    with $p(\,\cdot\,)$ of the form~\eqref{eq:true-joint} is $q_{\paramPD^*}(Z_k, X_k)$, where
    \begin{align} \label{eq:EPD-general-expression}
        \paramPD^*( \mathbf{x}_k ) &= \mathrm{E}_{ \RFSTheta_k | \mathbf{x}_k } \big[ \, P_{\ObjectGenRFS}(\mathbf{x}_k, \, O_k ) \, \big] 
    \end{align}
    is the \expectedProbabilityOfDetection{}. 
    The expectation is taken over the RPD~\eqref{eq:RPD-RFS-definition}. 
    Proof is given in Appendix~\ref{app:proof-of-proposition-KLD}.
\end{proposition}

Using $P_\ObjectGenRFS(\mathbf{x}_k) {\coloneqq} \paramPD^*( \mathbf{x}_k )$~\eqref{eq:EPD-general-expression} for the SPO model, we conclude that the density $ p_{ \mathrm{A}\text{-}\mathrm{SPO}_k } ( Z_k | X_k,\, \paramPD^* )$ best matches $p_{ \mathrm{A}\text{-}\mathrm{SPO}\text{-}\mathrm{D}_k } ( Z_k | X_k )$ for any $X_k$ distributed according to the prior density.
By considering the auxiliary likelihoods without marks using \eqref{eq:demarking-a-density}, we obtain the best SPO approximation for the intractable SPO-D model.

\section{SPO Approximation with Best \probabilityOfDetection{}: Marks 
Only}\label{sec:special-case}
As indicated before, implementations using GMs often assume a constant \probabilityOfDetection{}.
To address occlusions, however, the \probabilityOfDetection{} cannot be the same constant for all objects.
In the following, we deal with an efficient \emph{mark-only} case of Proposition~\ref{prop:best-fitting-KLD} to find a suitable constant for each object.
Assume that $\paramPD$ in~\eqref{eq:joint-with-parameter} is a function of only the mark, i.e., $\paramPD : \spaceOfMarks {\rightarrow} [0,1]$.
Redefining $Q$~\eqref{eq:space-of-functions-with-parameter} as
\begin{align}
    \! Q \triangleq \big\{ q_\paramPD(\,\cdot\,,\,\cdot\,) \text{ of the form~\eqref{eq:joint-with-parameter}} :\hspace{0.2cm} \paramPD : \spaceOfMarks {\rightarrow} [0,1] \big\} , \!
    \label{eq:space-of-functions-with-parameter-redefined}
\end{align}
and using it with~\eqref{eq:KLD-optimality} yields the optimal 
\begin{subequations}\label{eq:EPD-mark-case}
\begin{align}
     \paramPD^*( m ) &=  \mathrm{E}_{O_k,x_k|(\,\cdot\,,m)} \big[ \, P_\ObjectGenRFS\big((x_k,m)), O_k\big) \, \big] \\
    &\hspace{-0.0cm} \equiv \textstyle \iint P_\ObjectGenRFS\big((x_k,m)), O_k\big) \times \notag \\[-0.1cm]
    & \hspace{0.75cm} p_{\RFSTheta_k | \mathbf{x_k}}\big( O_k | (x_k,m) \big) D(x_k | m) \delta O_k \d x_k 
     , \!
\end{align}
\end{subequations}
where the single-object conditional and marginal densities for the given mark $m {\in} \spaceOfMarks$ are given by the PHD as 
\begin{align}
    \! \textstyle D(x_k | m) =  \tfrac{ D_{\RFSXi}(x_k,m) }{ D(m) }\,, \hspace{0.2cm} D(m) = \int D_{\RFSXi}(x_k,m) \d x_k \, . \!
    \label{eq:PHD-decomposition-xk,m}
\end{align}
As a special case of Proposition~\ref{prop:best-fitting-KLD}, the result~\eqref{eq:EPD-mark-case} provides the best-fitting \probabilityOfDetection{} as a function of only the mark.
Its practical use is explained in the following, while the proof is given in Appendix~\ref{app:proof-of-proposition-KLD} (Corollary~\ref{corollary-special-case-of-proposition}).

Recall that marks in $X_k$ must be distinct so that the auxiliary SPO and SPO-D likelihoods are A-PMB densities. 
It is apparent that this assumption is unnecessary for computing~\eqref{eq:EPD-mark-case}, while $X_k$ does not even need to be marked for the calculation of~\eqref{eq:EPD-general-expression}.
For such a case, however, the optimality (Proposition~\ref{prop:best-fitting-KLD} and thus Corollary~\ref{corollary-special-case-of-proposition}) was not established in this paper. 
To give a provably optimal end-to-end solution below, we remain consistent with the assumption on $X_k$ and use MBM filter for convenience. 

\subsection{Multi-Bernoulli Mixtures in MOT}
Assuming the birth is MB, the MBM density is a conjugate prior for the SPO model~\ref{sec:MOT-without-occlusions}, giving rise to the MBM filter~\cite{GM-MBM:2019}.
We equip each Bernoulli with a unique mark to compute the \expectedProbabilityOfDetection{} for each mark instead of each state.
Following Section~\ref{sec:auxiliary-variables} directly, the marks are called auxiliary variables in this paper and they are equivalent to labels.
In particular, we deal with the special case of A-PMB densities where the Poisson part is missing, which are addressed as auxiliary MB (A-MB) densities.
Note that the resulting algorithm is a special case of the MBM filter that is equivalent to the LMBM filter. 
For notational simplicity, the time index $k$ is dropped in the following.

\subsubsection{Auxiliary MB RFS}
The A-MB RFS results by dropping the Poisson component from the \mbox{A-PMB} RFS (Section~\ref{sec:auxiliary-variables}), and it can be regarded as the LMB RFS~\cite[pp. 453-458]{Mahler-Book:2014}.
The corresponding density can be written as
\begin{align}
    p_{\mathrm{A}\text{-}\mathrm{MB}} (X) &= \textstyle
    1_{ M } ( \mathcal{L}(X) ) \cdot \prod_{ \ell = 1 }^N p_{\mathrm{A}\text{-}\mathrm{Ber}}^{\mu^\ell} \big( X^{\mu^\ell} \big) \, .
\end{align}
If $m{\in} M {=}\{\mu^1,\dots,\mu^N\}$, its RPD is a special case of~\eqref{eq:RPD-for-standard-PMB-density}, 
\begin{align}
    p_{ \mathrm{A}\text{-}\mathrm{MB:} \RFSTheta | \mathbf{x}}(O | (x,m)) &= 
    p_{\mathrm{A}\text{-}\mathrm{MB} \setminus m}( O ) \, ,
\end{align}
and zero otherwise, with $p_{\mathrm{A}\text{-}\mathrm{MB} \setminus m}( O )$ given in~\eqref{eq:MB-without-ell}

\subsubsection{Auxiliary MBM RFS}
A mixture of A-MB-distributed RFSs is referred to as the auxiliary MBM (A-MBM) RFS, which can be viewed as the LMBM RFS~\cite{Mahler-integralTransform:2016}.
In MOT, each mixture term corresponds to a \emph{global data association hypothesis}.
Assigning a subindex $h$ to each term/hypothesis, the density of the \mbox{A-MBM} RFS becomes
\begin{align}
    p_{\mathrm{A}\text{-}\mathrm{MBM}} ( X ) &= \textstyle \sum_{h=1}^{\mathcal{H}} w_h \cdot p_{\mathrm{A}\text{-}\mathrm{MB}, h}(X) \, ,
    \label{eq:LMBM-density}
\end{align}
where $\mathcal{H}$ is the number of mixture terms, and $w_h$ is the weight of the hypothesis $h$, $h {=} 1,\dots,\mathcal{H}$, such that $\sum_{h=1}^{\mathcal{H}} w_h = 1$.
Let $M_h {\triangleq} \{\mu_h^\ell\}_{\ell=1}^{N_h}$ be the set of auxiliary variables forming the support corresponding to the A-MB hypothesis $h$, and extend the existence probability $r_h^m$ to be zero if $m {\notin} M_h$.
The RPD for the A-MBM RFS is then
\begin{align}
    & p_{ \mathrm{A}\text{-}\mathrm{MBM:} \RFSTheta | \mathbf{x}} \big(O | (x,m)\big) \\[-0.1cm]
    & \hspace{0.0cm} \textstyle = \tfrac{1}{D_{\mathrm{A}\text{-}\mathrm{MBM}}(x,m)} \! \sum_{h=1}^{\mathcal{H}} \! w_h {\cdot} r_h^m {\cdot} p_{\mathrm{sp},h}(x|m) {\cdot} p_{\mathrm{A}\text{-}\mathrm{MB},h \setminus m} (O) , \!\notag
\end{align}
where the PHD of the A-MBM RFS is 
\begin{align}
    D_{\mathrm{A}\text{-}\mathrm{MBM}}(x,m) &= \textstyle \sum_{h=1}^{\mathcal{H}} w_h {\cdot} r_h^{m} {\cdot} p_{\mathrm{sp},h} (x|m) \, . 
\end{align}
Employing the PHD conditioning~\eqref{eq:PHD-decomposition-xk,m} yields 
\begin{subequations}
\begin{align}
    D_{\mathrm{A}\text{-}\mathrm{MBM}} (x | m) &= \tfrac{ \sum_{h=1}^{\mathcal{H}} w_h {\cdot} r_h^{m} {\cdot} p_{\mathrm{sp},h} (x|m) }{ D_{\mathrm{A}\text{-}\mathrm{MBM}} ( m ) } \, , \\[-0.1cm]
    D_{\mathrm{A}\text{-}\mathrm{MBM}} ( m ) &= \textstyle \sum_{h=1}^{\mathcal{H}} w_h {\cdot} r_h^{m} \, .
\end{align}
\end{subequations}
The conditional density for the expectation~\eqref{eq:EPD-mark-case} becomes
\begin{align}
    & p_{ \mathrm{A}\text{-}\mathrm{MBM:} \RFSTheta | \mathbf{x}} \big( O | (x,m) \big) D_{\mathrm{A}\text{-}\mathrm{MBM}} (x | m)  \\[-0.1cm]
    & \hspace{0.0cm} \textstyle = \tfrac{1}{D_{\mathrm{A}\text{-}\mathrm{MBM}}(m)} \! \sum_{h=1}^{\mathcal{H}} \! w_h {\cdot} r_h^m {\cdot} p_{\mathrm{sp},h}(x|m) {\cdot} p_{\mathrm{A}\text{-}\mathrm{MB},h \setminus m} (O) \, . \notag
\end{align}

\subsection{Optimal parameter for MBM Filtering}
Once the A-MBM prior density is obtained in the form~\eqref{eq:LMBM-density}, the best-fitting approximation~\eqref{eq:EPD-mark-case} becomes
\begin{align}
    P_{\ObjectGenRFS} (m) \coloneqq \textstyle \tfrac{1}{ \sum_{h=1}^{\mathcal{H}} w_h {\cdot} r_h^{m} } \sum_{h=1}^{\mathcal{H}} w_h {\cdot} r_h^{m} {\cdot} \bar{P}_{\ObjectGenRFS}(m,h) \,, \label{eq:Pd-m-only}
\end{align}
where the \expectedProbabilityOfDetection{} corresponding to the hypothesis $h$ is
\begin{align}
    & \bar{P}_{\ObjectGenRFS}(m,h) \triangleq \notag \\[-0.1cm]
    & \hspace{0.0cm}\textstyle \iint P_{\ObjectGenRFS}\big((x,m), O\big) \,
    p_{\mathrm{A}\text{-}\mathrm{MB},h \setminus m}( O ) p_{\mathrm{sp},h}(x|m) \, \delta O \d x .
    \label{eq:EPD-special-case-computation}
\end{align}
Since $P_\ObjectGenRFS(\,\cdot\,,\,\cdot\,)$ cannot generally be written as a sum or a product, the computation of~\eqref{eq:EPD-special-case-computation} is challenging.
We propose to use directly the set integral form of~\eqref{eq:EPD-special-case-computation}, i.e., 
\begin{align}
    &\hspace{-0.0cm} \bar{P}_{\ObjectGenRFS}(m,h) = \textstyle \sum_{n=0}^{|M_h\setminus \{m\}|} 
    \sum_{ A \subseteq (M_h \setminus \{m\} ), \text{ s.t. } |A|=n } \omega_h^m(A)
    \notag \\[-0.1cm]
    & \hspace{0cm} \textstyle \int \underbrace{ \textstyle \int\cdots\int }_{n \text{-times} } P_{\ObjectGenRFS}\big( (x,m), \{ (o^\mu, \mu )\}_{\mu \in A} \big) \times 
    \notag \\[-0.6cm]
    & \hspace{1.7cm} \textstyle p_{\mathrm{sp},h}(x|m) \big( \prod_{\mu \in A}  p_{\mathrm{sp},h}(o^\mu | \mu) \big) \d x \prod_{\mu \in A} \d o^\mu \, ,
    \label{eq:EPD-special-case-computation:set-integral}
\end{align}
if $m {\in} M_h$ and zero otherwise, where 
\begin{align}
    \omega_h^m(A) \triangleq \textstyle \big( \prod_{\mu \in (M_h \setminus \{m\} ) \setminus A } (1-r_h^\mu) \big) \big( \prod_{\mu \in A} r_h^\mu \big) \, , \label{eq:LMB-set-function-omega}
\end{align}
is a set function, c.f.~\cite[p.~454]{Mahler-Book:2014}.
Note that~\eqref{eq:EPD-special-case-computation:set-integral} generally has $\sum_{n=0}^{|M_h|-1} \left( \begin{smallmatrix} |M_h|{-}1 \\ n \end{smallmatrix} \right) = 2^{|M_h|-1}$ terms and the complexity of their evaluation ranges greatly among them.

In practice, the number of other objects with high existence probability that are likely to occlude the object with mark $m$ is often rather small.
In such a case, the sums in~\eqref{eq:EPD-special-case-computation:set-integral} involve many insignificant terms. 
Next, we propose a reduction and efficient approximation of~\eqref{eq:EPD-special-case-computation:set-integral}.

\subsubsection{Simplification of the Hypothesis $h$}\label{subsubsec:simplify-LMB}
First, reduce the set $M_h {\setminus} \{m\}$ by discarding Bernoulli components with insignificant existence probabilities.
Further reduction can be based on the following axiom.
\begin{axiom}[Objects occlude individually]\label{axiom:objects-occlude-individually}
    \probabilityOfDetection{} is a locally integrable function satisfying the implication
    \begin{align}
        \!\!\! P_{\ObjectGenRFS}\big( \mathbf{x}, \! \{\mathbf{o}\} \big) {=} P_{\ObjectGenRFS}( \mathbf{x}, \emptyset )  \Rightarrow P_{\ObjectGenRFS}\big( \mathbf{x}, O {\cup} \{\mathbf{o}\} \big) {=} P_{\ObjectGenRFS}( \mathbf{x}, \! O ) , \!\!\!
    \end{align}
    for any $\mathbf{x},\mathbf{o} {\in} \X$ and any finite $O {\subset} \X$. 
    In essence, if an object $\mathbf{o}$ does not influence the \probabilityOfDetection{} of the object $\mathbf{x}$ (i.e., does not occlude $\mathbf{x}$) on its own, then it never will, regardless of the presence of any other object.\\[-0.5cm]
\end{axiom}

\begin{corollary}\label{corollary:objects-occlude-individually}
    Let $\mathbf{x}$ and $\mathbf{o}$ be random variables on $\X$ with continuous (spatial) density functions, then
    \begin{align}
        & \textstyle \mathrm{E}_{\mathbf{x},\mathbf{o}} \big[ P_{\ObjectGenRFS}\big( \mathbf{x}, \! \{\mathbf{o}\} \big) \big] {=} \mathrm{E}_{\mathbf{x}} \big[ P_{\ObjectGenRFS}( \mathbf{x}, \emptyset ) \big] \notag \\
        & \hspace{1cm} \Rightarrow \ \mathrm{E}_{\mathbf{x},\mathbf{o}} \big[ P_{\ObjectGenRFS}\big( \mathbf{x}, O {\cup} \{\mathbf{o}\} \big) \big] {=} \mathrm{E}_{\mathbf{x}} \big[ P_{\ObjectGenRFS}( \mathbf{x}, \! O ) \big] \, ,
    \end{align}
    for any finite $O {\subset} \X$, where $\X$ is further assumed to be \emph{directionally limited}\footnote{
        This geometric assumption is key to the Lebesgue differentiation theorem used in the proof.
        Spaces like $\mathbb{R}^n$ satisfy this condition automatically.
    }~\cite[p.~7]{Heinonen:AnalysisMetricSpaces:2001}.
    In words, if an object $\mathbf{o}$ does not affect the integral of the \probabilityOfDetection{} of $\mathbf{x}$ itself, it will not do so in the presence of other objects either.
    The proof is given in Appendix~\ref{app:proof-of-corollary:objects-occlude-individually}.\\[-0.5cm]
\end{corollary}

It follows that one can discard $\mu {\in} M_h {\setminus} \{m\}$, for which
\newcommand{\PdAlnoe}{P_{\ObjectGenRFS}^{\emptyset}(m,h)}
\begin{align}
    \mathrm{E}_{x,o} \big[ P_{\ObjectGenRFS}\big( (x,m), \! \{ (o,\mu) \} \big) \big] \approx \underbrace{ \mathrm{E}_{x} \big[ P_{\ObjectGenRFS}( (x,m), \emptyset ) \big] }_{ \triangleq \PdAlnoe } \,,
    \label{eq:pD-m-alone}
\end{align}
where the densities used to compute the expectations are taken from the hypothesis $h$.
Both the left and right-hand sides of~\eqref{eq:pD-m-alone} can be implemented using Monte Carlo (MC) integration.
From an implementation perspective, the MC samples should be stored for future reference.

\emph{Remark:}
In the absence of any occlusions among all hypotheses, $\bar{P}_{\ObjectGenRFS}(m,h) {=} P_D^{\emptyset}(m,h)$~\eqref{eq:pD-m-alone}.
The complexity of the \expectedProbabilityOfDetection{} computation is then 
$\mathcal{O}( \, \mathcal{H} M_{\max}^2 N_{\mathrm{MC}} \, )$
in the worst\discretionary{-}{-}{-}case, where $M_{\max} = \max \{|M_1|,\dots,|M_{\mathcal{H}}|\}$ and $N_{\mathrm{MC}}$ is the number of MC samples.

Finally, we approximate existence probabilities that can be considered high enough by one, so that any combination-set $A$ in~\eqref{eq:EPD-special-case-computation:set-integral} excluding the corresponding marks leads to negligible $\omega_h^m(A)$~\eqref{eq:LMB-set-function-omega}.

Note that this simplification can be computed in a track-oriented manner.
That is, for each \emph{local} hypothesis corresponding to the mark $m$, one can go through local hypotheses of other objects included in the same global hypothesis.
The simplification for the given $m$ is likely to be the same among many global hypotheses, eventually leading to the same approximation of~\eqref{eq:EPD-special-case-computation:set-integral} for them.

For the next approximation step, it would be generally possible to employ the \mbox{$K$-shortest} path algorithm used in the early implementations of the GLMB filter~\cite{VoVo-dGLMB:2014}.
However, only a small number of other Bernoullies in the set $M_h {\setminus} \{m\}$ are likely to remain for consideration after the simplification~\ref{subsubsec:simplify-LMB}.
Therefore, we find the following exhaustive computation readily suited. 

\subsubsection{Approximation of Eq.~\eqref{eq:EPD-special-case-computation:set-integral} Directly}\label{subsubsec:approximate-PDmh}
Denote with $M_h^c {\subset} M_h {\setminus} \{m\}$ the subset of \emph{certain} marks for which existence probabilities are equal to one, and denote its cardinality with $n^c {=} |M_h^c|$.
Similarly, denote with $M_h^u {=} M_h {\setminus} \{m\} {\setminus} M_h^c$ the set of \emph{uncertain} marks, and denote its cardinality with $n^u {=} |M_h^u|$.
For each $n {=} n^c,n^c{+}1,\dots,n^c{+}n^u$ compute the terms of the outmost sum in~\eqref{eq:EPD-special-case-computation:set-integral} as follows.
First, generate directly the subsets (i.e., combinations) $A^u {\subseteq} M_h^{u}, $ with $ |A^u| {=} \nu$ for each $\nu {=} 0, \dots, n^u $, such that $A {=} A^u {\cup}  M_h^c$.
For each such $A$, approximate the corresponding $(n^c{+}\nu{+}1)$-way integral in~\eqref{eq:EPD-special-case-computation:set-integral} with, e.g., the MC integration. 

\begin{figure}
    \centering
    \subfloat[Experimental cost.]{
        \includegraphics[width=0.49\linewidth,trim={0.1cm, 1.7cm, 0.1cm, 2.1cm},clip]{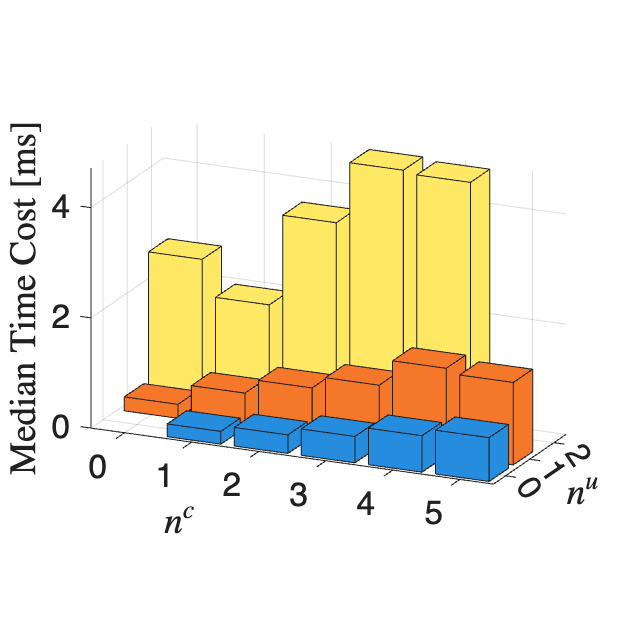}
    }
    \hspace{-0.5cm}
    \subfloat[Theoretical cost.]{
        \includegraphics[width=0.49\linewidth,trim={0.1cm, 1.5cm, 0.1cm, 2.3cm},clip]{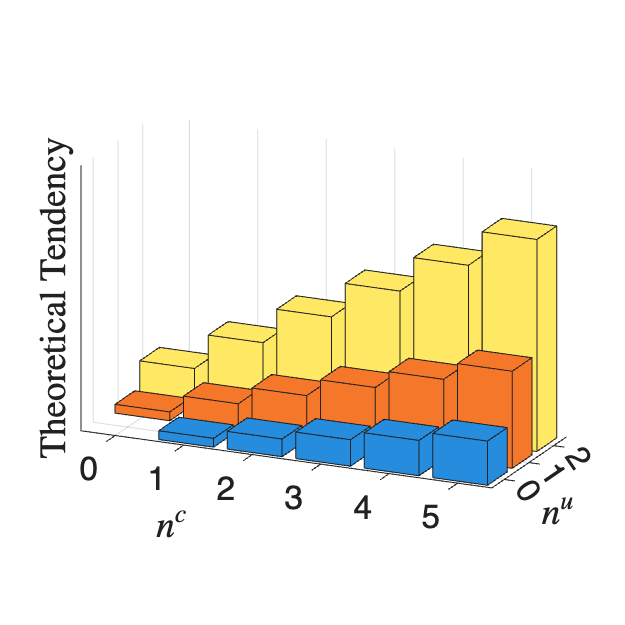}
    }
    \vspace{-0.1cm}
    \caption{
        Complexity analysis of the approximation~\ref{subsubsec:approximate-PDmh}.
        Not enough data for $n^c{=}5$ and $n^u{=}2$ in (a) were available.
    }
    \label{fig:Complexity-analysis}
    \vspace{-3mm}
\end{figure}

Under the simplification~\ref{subsubsec:simplify-LMB}, the total number $2^{|M_h|-1}$ of terms in~\eqref{eq:EPD-special-case-computation:set-integral} shrinks to only $2^{(n^u)}$.
Neglecting the complexity of evaluating $\omega_h^m(A)$~\eqref{eq:LMB-set-function-omega} and assuming the complexity of evaluating an $(n^c {+} \nu {+} 1)$\discretionary{-}{-}{-}way integral in~\eqref{eq:EPD-special-case-computation:set-integral} is roughly linear in both $n^c$ and $n^u$, the computational complexity of the approximation~\ref{subsubsec:approximate-PDmh} is $\mathcal{O}\big( N_{MC} (n^c {+} n^u) 2^{(n^u)} \big)$.
The theoretical dependency for small $n^c$ and $n^u$ is compared with experimental time costs (median values) in Fig.~\ref{fig:Complexity-analysis}.
Details about the experiment are given later in Section~\ref{sec:Results}.

\subsection{Summary of the Proposed Strategy}
The proposed filter utilizes the following strategy to deal with occlusions.
To distinguish it from the \solVo{} strategy (Section~\ref{sec:accounting-for-occlusions}), it is termed as the \ourName{} strategy: 
\begin{itemize}
    \item[\ourOne\,:] Palm condition the prior distribution of states, 
    \item[\ourTwo:] reduce the dependency within the SPO-D model by computing the EPD $P_{\ObjectGenRFS}(\,\cdot\,)$,
    \item[\ourThree:] obey the SPO model parameterized with the computed $P_{\ObjectGenRFS}(\,\cdot\,)$ to achieve tractable filtering recursion.
\end{itemize}

The proposed filter will be referred to as MBM-\ourName{}. 
Analogically, if the \solVo{} strategy (Section~\ref{sec:accounting-for-occlusions}) is applied instead, the algorithm will be referred to as MBM-\solVo{}. 



\newcommand{\elementOfvector}[2]{[#1]_{#2}}
\newcommand{\area}{\text{Area}}
\newcommand{\BB}{\text{BB}}
\section{Application to Visual MOT}
\label{sec:Results}

\begin{figure}[t]
	\centering
        \vspace{-2mm} 
    \includegraphics[width=0.63\linewidth]{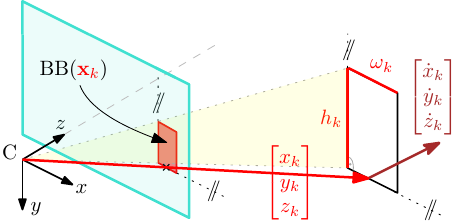}
	\caption{
    Illustration of the state $\mathbf{x}_k$~\eqref{eq:state-3D-pom} in red.
    The image plane is in turquoise, and $\mathrm{C}$ is the camera center.
    }\label{fig:camera_geometry}
        \vspace{-5mm} 
\end{figure}

In~\cite{KrKoXiLeSt:SPO:2025}, an SPO model for pedestrian MOT using a monocular camera was developed with parameters obtained by identification and physical modeling.
The MOT-17 dataset~\cite{MOT-16:2016,Dendorfer:MOTChallenge:2021} was considered, and the Faster R-CNN (FRCNN) bounding box (BB) detections included in the dataset were used for the identification.
Pedestrians were modeled in 3D with the state at time step $k$ defined as~\cite{KrKoStDu:2024:IEEE}
\begin{align}
    \mathbf{x}_k = [ \xThreeDee_k\ \xVelocityThreeDee_k\ \yThreeDee_k\ \yVelocityThreeDee_k\ \zThreeDee_k\ \zVelocityThreeDee_k\ \widthThreeDee_k\ \heightThreeDee_k ]\T,\label{eq:state-3D-pom}
\end{align}
with the position $\xThreeDee_k$, $\yThreeDee_k$, $\zThreeDee_k$ of the lower-bottom center of the box and the width $\widthThreeDee_k$ and height $\heightThreeDee_k$ expressed in meters, while the velocities $\xVelocityThreeDee_k$, $\yVelocityThreeDee_k$, $\zVelocityThreeDee_k$ in meters per second.
Geometric illustration of the measurement and the state quantities is given in Fig.~\ref{fig:camera_geometry}.
For details, please refer to~\cite{KrKoStDu:2024:IEEE}.
The fifth state-vector element $\elementOfvector{\mathbf{x}_k}{5} {=} z_k$ will be called \emph{depth} in the sequel.


The SPO model~\cite{KrKoXiLeSt:SPO:2025} involves the constant $P_D{=}0.529$, which was estimated from the entire MOT-17 training dataset. 
Such a modeling was proven to be conceptually incorrect in~\cite{KrKoXiLeSt:SPO:2025}, where the \probabilityOfDetection{} was shown to depend on the \emph{visibility} as follows. 
\begin{minipage}{\linewidth}
    \vspace{0.1cm}
    \begin{wrapfigure}[8]{l}{0.39\textwidth}
        \vspace{-3.5mm}
        \begin{center}
        \includegraphics[width=1\linewidth]{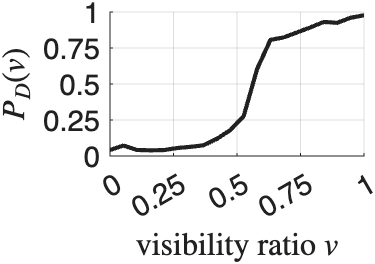}
        \vspace{-0.6cm}
        \caption{
        \probabilityOfDetection{} as a function of visibility ratio. 
        }
        \label{fig:P_D(v):MOT-17}
        \end{center}
    \end{wrapfigure}
    In~\cite{KrKoXiLeSt:SPO:2025}, an estimate $P_D(v)$ of the \probabilityOfDetection{} was computed \emph{conditionally} on the \emph{visibility ratio} $v {\in} [0,1]$, where $v$ is available in ground-truth (\groundTruth{}) data~\cite[Table~5]{MOT-16:2016}.
    The resulting $P_D(v)$ curve is far from being constant and is shown in Fig.~\ref{fig:P_D(v):MOT-17}. 
    \vspace{0.12cm}
\end{minipage}

\emph{Remark:}
The true \probabilityOfDetection{} likely depends on conditions such as brightness or resolution. 
The $P_D(v)$ (Fig.~\ref{fig:P_D(v):MOT-17}) is an average over such phenomenons contained in the data.
    
\emph{Remark:} the ratio $v$ from~\cite{MOT-16:2016} capture the visibility of pedestrian \emph{BBs} rather than of their \emph{bodies}.
Physically, the \probabilityOfDetection{} should depend on the latter visibility.
To follow such a modeling, however, a more informative state than $\mathbf{x}_k$~\eqref{eq:state-3D-pom} is required.
Accordingly, image segmentation measurements could be better suited for such consideration than BB detections used here.
Finally, establishing the corresponding \probabilityOfDetection{} model may be challenging.
Therefore, the former (i.e., BB) \probabilityOfDetection{} visibility dependence modeling is adopted in this paper.
While \emph{heuristic} models for such dependency exist~\cite{Occlusions-GLMB-Vo:2022,Linh:VisualOccluLMB:2024,OccluGLMB:multiView:2024}, the function $P_D( v )$ (Fig.~\ref{fig:P_D(v):MOT-17}) \emph{estimated from data} is used in this paper.

\begin{figure}[h]
    \centering
    \includegraphics[scale=0.76]{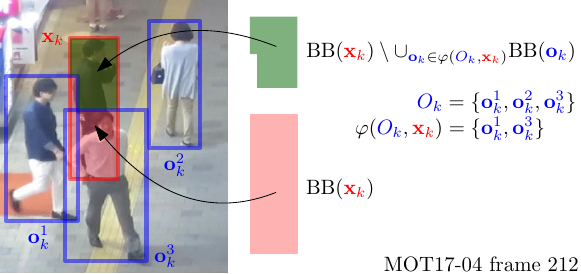}
    \caption{
    Illustration of sets whose area is computed in~\eqref{eq:visibility-ratio}.
    }
    \label{fig:illustration_of_occlusions}
    \vspace{-5mm} 
\end{figure}

The BB visibility ratio $v$ definition was rather brief in~\cite{MOT-16:2016}.
In this paper, the ratio $v$ is modeled without any reference to ground plane, c.f.~\cite[Sec.~2.5]{MOT-16:2016}.
In particular, a definition similar to~\cite[Sec.~4.2]{OccluGLMB:multiView:2024} (c.f.~\cite{MonocularMTT3dOcclu:2013}) is used as
\begin{align}
    \!\! v( \mathbf{x}_k, \, O_k ) &= \frac{ \area{}\big( \BB( \mathbf{x}_k ) {\setminus} \cup_{\mathbf{o}_k \in \varphi(O_k, \mathbf{x}_k) } \! \BB(\mathbf{o}_k) \big) }{ \area{}\big( \BB( \mathbf{x}_k )  \big) } , \!\! 
    \label{eq:visibility-ratio}
\end{align}
where the function $\BB(\,\cdot\,) {\subset} \mathbb{R}^2$ yields the bounding box corresponding to the state as illustrated in Fig.~\ref{fig:illustration_of_occlusions}, and 
\begin{align}
    \! \varphi(O_k, \mathbf{x}_k) {=} \{ \mathbf{o}_k {\in} O_k \! : 
    \elementOfvector{\mathbf{o}_k}{5} {<} z_{\max} , \,
    \elementOfvector{\mathbf{o}_k}{5} {<} \elementOfvector{\mathbf{x}_k}{5} {-} \kappa \} \!
\end{align}
is the subset of other objects that are considered eligible to occlude $\mathbf{x}_k$.
The condition $\elementOfvector{\mathbf{o}_k}{5} {<} z_{\max}$ means that occlusions by objects that appear further than $z_{\max} {=} 15$ meters are ignored due to limited depth estimation accuracy for distant objects.
The $z_{\max}$ value was selected to mitigate meaningless EPoD values for crowded pedestrians in the distance, and experiments exhibited rather small sensitivity to the value.
The condition $\elementOfvector{\mathbf{o}_k}{5} {<} \elementOfvector{\mathbf{x}_k}{5} {-} \Delta$ means that occlusions by objects that are in front of $\mathbf{x}_k$ but too close in $z$-coordinate are ignored to ensure physical validity of occlusion.
Here, $\kappa {=} \tfrac{0.85}{2}$ meters was set to half of a mean BB width~\cite{KrKoXiLeSt:SPO:2025}.
Experiments validated that using either a smaller or larger $\kappa$ may lead to meaningless results.

Assuming the \probabilityOfDetection{} depends on pedestrians via the visibility ratio \emph{only}, 
it follows that
\begin{align}
    P_D( \mathbf{x}_k,\, O_k) &= P_D\big( \, v( \mathbf{x}_k, \, O_k ) \, \big) \, ,
    \label{eq:PD_v}
\end{align}
where $P_D(\,\cdot\,)$ on the right-hand side is $P_D(v)$ (Fig.~\ref{fig:P_D(v):MOT-17}).

The SPO model~\cite{KrKoXiLeSt:SPO:2025} is adopted with the approximate GM implementation~\cite{KrKoXiLeSt:SPO:2025}.
However, the methods developed in this paper rely on the assumption that the birth density is MB, which is adopted based on~\cite[Sec.~II.C.1]{KrKoXiLeSt:SPO:2025}.
A suitable MB is also used as the initial density.
Available FRCNN detections were used as the measurements.

For each time step, estimates are drawn for simplicity using the sub-optimal Estimator 1~\cite[Sec.~VI.A]{PMBM:2018} with a \emph{non-informative} threshold of $0.5$ for existence probabilities.
Since only 2D \groundTruth{} data are available, estimates are further projected into 2D for evaluation, which is achieved using the Unscented transform as in~\cite[Sec.~IV]{KrKoStDu:2024:IEEE}. 
Ellipsoidal gating was used with a threshold of 6. 
The number of global hypotheses was capped at 100, with a pruning threshold of -300 for log-weights.
For hypothesis $h$, the Murty's $M$-best assignments algorithm was used with $M {=} \lceil w_h \cdot 10 \rceil$~\cite{VoVo-dGLMB:2014,PMBM:2018}.

\subsection{Performance Evaluation Metric}
To evaluate the estimates, CV scores such as \emph{multiple object tracking accuracy} (MOTA), \emph{higher order tracking accuracy} (HOTA), and \emph{identity F1} (IDF1) are often considered in the CV literature~\cite{MOT-16:2016,HOTA:2021}. 
However, the CV scores may behave undesirably~\cite{Nguyen:Visual-metrics-trustworthy:2023,KrKoStXiSvFe:TGOSPA-params-visual:ArXiv} and hardly allow for performance evaluation based on user preferences. 
Therefore, we employ the \emph{trajectory generalized sub-optimal assignment} (TGOSPA) metric~\cite{TrajectoryGOSPA:2020}, which includes several parameters: (i)~metric $d(x,y)$ between GT BB $x$ and an estimated BB $y$, (ii)~power parameter $p {\geq} 1$, (iii)~cut-off $c {>} 0$ that is the maximum possible distance for $x$ and $y$ to be assigned to each other, and (iv)~switching penalty $\gamma {>} 0$.
The parameters allow the TGOSPA metric to be tailored for applications such as online surveillance or offline scene understanding~\cite{KrKoStXiSvFe:TGOSPA-params-visual:ArXiv}. 

\newcommand{\Err}{\mathsf{E}}
\newcommand{\No}{\mathsf{N}}
The TGOSPA metric value can be decomposed into errors $\Err$ resulting from \emph{(i: }$\Err_{\text{TP}}$\emph{)} properly estimated objects, i.e., \emph{true positives} (TP), \emph{(ii: }$\Err_{\text{FN}}$\emph{)} missed objects, i.e., \emph{false negatives} (FN), \emph{(iii: }$\Err_{\text{FP}}$\emph{)} false objects, i.e., \emph{false positives} (FP), and \emph{(iv: }$\Err_{\text{Sw}}$\emph{)} properly defined \emph{track switches} (Sw).
Informally, the TGOSPA metric is given by
\begin{align}
    \label{eq:TGOSPA-informally}
    & d_{p}^{(c,\gamma)}( \mathbf{X}, \mathbf{Y}) = \textstyle \min_{ \text{``over assignments between trajectories''} } \\[-0.1cm]
    & \Big( \, \underbrace{ \textstyle \sum_{(x,y) \in \text{TP} } d(x,y)^p }_{ \Err_{\text{TP}} } 
    + \underbrace{ \tfrac{c^p}{2} |\text{FN}| }_{ \Err_{\text{FN}} }
    + \underbrace{ \tfrac{c^p}{2} |\text{FP}| }_{ \Err_{\text{FP}} }
    + \underbrace{ \gamma^p \text{Sw} }_{ \Err_{\text{Sw}} } \, \Big)^{\sfrac{1}{p}} , \notag
\end{align}
where $\mathbf{X}$ and $\mathbf{Y}$ are sets of \groundTruth{} and estimated trajectories, respectively, the symbol $|\cdot|$ denotes the cardinality, and Sw counts changes within the assignments as
\begin{align}
    \text{Sw} = \textstyle \sum_{k=1}^K \sum_{i = 1}^{|X|}
    \underbrace{
        s(\text{assignment}_{k-1}^i,\ \text{assignment}_{k}^i)
    }_{
        \footnotesize
        \hspace{-2.0cm}
        \triangleq \begin{cases}
            0 & \text{if assignment}_{k-1}^i = \text{assignment}_{k}^i \, , \\[-0.1cm]
            1 & \text{if } 0 \neq \text{assignment}_{k-1}^i \neq \text{assignment}_{k-1}^i \neq 0 \, ,  \\[-0.1cm]
            \tfrac{1}{2}  & \text{otherwise},
        \end{cases}
    } 
\end{align}
where assignment$_k^i$ is the index $j {\in} \{0,1,\dots,|Y|\}$ to which the $i$-th \groundTruth{} trajectory is assigned to at time step~$k$ with~$0$ meaning it is left unassigned.
Note that all of TP, FP, and FN depend on the assignments as well.
See~\cite{TrajectoryGOSPA:2020} for a precise definition and approximate computation of~\eqref{eq:TGOSPA-informally}.

The desired meaning and impact of the diffident terms in~\eqref{eq:TGOSPA-informally} vary among applications. 
In the following, the guidelines presented in~\cite{KrKoStXiSvFe:TGOSPA-params-visual:ArXiv} are used to select $d$, $p$, $c$ and $\gamma$ concisely.
The terms  $\Err_\text{TP}$, $\Err_\text{FN}$ and $\Err_\text{FP}$ are further decomposed to better assess performance under occlusions.

\subsubsection{TGOSPA Metric For Assessing Occlusions}
The function $d(x,y) = 1 {-} \mathrm{IoU}(x,y)$ with $\mathrm{IoU}$ being the \emph{intersection over union} was proven to be a metric in~\cite{KrKoStXiSvFe:TGOSPA-params-visual:ArXiv} and it is adopted in this paper.
The maximum possible cut-off $c {=} 1$ is used, allowing any two BBs to be assigned as long as they have a nonempty intersection.
The power parameter $p {=} 2.41$ is selected according to~\cite[Eq.~(32)]{KrKoStXiSvFe:TGOSPA-params-visual:ArXiv} so that estimates with error $0.75 {=} \tfrac{c}{\sqrt[p]{2}}$ in $d$ or larger are better to be omitted.
That is, tracking algorithms are encouraged to output even very distant predictive estimates with errors up to $0.75$ in the metric $d$.
Since $p{>}1$, the E$_\text{FP}$ and E$_\text{FN}$ can be expected to have the largest influence on the final metric value, compared to E$_\text{TP}$ and Sw.
The switching penalty $\gamma {=} 2.60 $ is set according to~\cite[Eq.~(32)]{KrKoStXiSvFe:TGOSPA-params-visual:ArXiv} so that switches Sw encapsulate long-term track changes that last for at least $10$ time steps\footnote{
    The videos considered in this paper were captured with $30$ frames per second, i.e., $10$ time steps correspond to $0.33$ seconds of a video.
} (see~\cite[Sec.~4.4.2]{KrKoStXiSvFe:TGOSPA-params-visual:ArXiv}).
Such a choice of $\gamma$ favors TP pairs that correspond to the same trajectories over pairs with the smallest error for a given time step.
Tracking algorithms are thus encouraged to form trajectories without fragments, such as those caused by occlusions.

The visibility ratio is available for each \groundTruth{} BB $x$ in the MOT-17 dataset, further denoted as $v_x$.
The error $\Err_{\text{TP}}$ can thus be decomposed as
\begin{align}
    &\Err_{\text{TP}} = \textstyle \sum_{(x,y) \in \text{TP} } (1{-}v_x + v_x ) d(x,y)^p \nonumber \\
    &= \underbrace{ \textstyle \sum_{(x,y) \in \text{TP} } (1{-}v_x) d(x,y)^p }_{ \Err_{\text{TP}}^{\text{o}} } + \underbrace{ \textstyle \sum_{(x,y) \in \text{TP} } v_x d(x,y)^p }_{ \Err_{\text{TP}}^{\text{v}} } \, , \notag\\[-0.5cm]
\end{align}
where $\Err_{\text{TP}}^{\text{o}}$ and $\Err_{\text{TP}}^{\text{v}}$ are the portions of $\Err_{\text{TP}}$ corresponding to cases when the \groundTruth{} BBs were occluded and visible, respectively.
Furthermore, 
\begin{align}
    |\text{TP}| 
    &= \underbrace{ \textstyle \sum_{(x,y) \in \text{TP} } (1{-}v_x) }_{ \No_{\text{TP}}^{\text{o}} } + \underbrace{ \textstyle \sum_{(x,y) \in \text{TP} } v_x }_{ \No_{\text{TP}}^{\text{v}} } \, ,
\end{align}
where $\No_{\text{TP}}^{\text{o}}$ and $\No_{\text{TP}}^{\text{v}}$ correspond to occluded and visible BBs, respectively. 
The error $\Err_{\text{FN}}$ can be decomposed as
\begin{align}
    \Err_{\text{FN}} &= \overbrace{ \tfrac{c^p}{2}  \underbrace{ \textstyle \sum_{x \in \text{FN} } (1{-}v_x) }_{ \No_{\text{FN}}^{\text{o}} } }^{ \Err_{\text{FN}}^{\text{o}} } + \overbrace{ \tfrac{c^p}{2} \underbrace{ \textstyle \sum_{x \in \text{FN} } v_x }_{ \No_{\text{FN}}^{\text{v}} } }^{ \Err_{\text{FN}}^{\text{v}} }
\end{align}
where $\Err_{\text{FN}}^{\text{o}}$ and $\Err_{\text{FN}}^{\text{v}}$ are the parts of $\Err_{\text{FN}}$ for the cases when missed \groundTruth{} BBs were occluded or visible, respectively.

\subsection{Results and Discussion}
To analyze the performance in challenging real scenarios, the scenarios should obey the assumed SPO-D model.
In particular, the camera must be static, and occlusions should be present. 
We thus selected MO17-02, MOT17-04, and MOT17-09 videos from the MOT-17 dataset. 

Performance is evaluated for the MBM filter with \emph{(i)}~the proposed~\ourName{} and \emph{(ii)}~the existing~\solVo{} (\cite{Occlusions-GLMB-Vo:2022,OccluGLMB:multiView:2024,Linh:VisualOccluLMB:2024}, see Section~\ref{sec:accounting-for-occlusions}) occlusion\discretionary{-}{-}{-}handling strategies that both utilize $P_D( \mathbf{x}_k, O_k)$~\eqref{eq:PD_v}, and also for \emph{(iii)}~the ``basic'' MBM filter with constant $P_D {=} 0.529$ as a reference.
Additionally, \emph{(iv)}~the SORT~\cite{SORT:2016} and \emph{(v)}~FasterTracker~\cite{FastTracker:2025} algorithms were considered as representatives of CV-based algorithms. The SORT algorithm processes FRCNN detections without using video images directly, similarly to the MBM filters, and it is evaluated for a \emph{fair} comparison with CV-based algorithms.
The FastTracker algorithm is evaluated as a representative of re-ID-based strategies.
It processes both the video images and the FRCNN detections and dominates the MOT-17 Public leader board\footnote{\url{motchallenge.net/results/MOT17/}}.
The results for \mbox{MOT17-02}, \mbox{MOT17-04} and \mbox{MOT17-09} videos are given in Tables~\ref{tab:MOT17-02}, \ref{tab:MOT17-04} and~\ref{tab:MOT17-09}, respectively.

\begin{table*}
    \centering
    \caption{Results for MOT17-02 Video}
    \vspace{-0.1cm}
    \label{tab:MOT17-02}
    \begin{tabular}{cccc|c|ccc|ccc|c|c|c}
    \toprule
            & \multicolumn{3}{c|}{CV scores ($\uparrow$)} & \multicolumn{9}{c|}{TGOSPA metric ($\downarrow$) and its decomposition} & Hz ($\uparrow$) \\[0.1cm]
            & MOTA & HOTA & IDF1 & $d_{p}^{(c,\gamma)}$ & $\substack{ \Err_{\text{TP}} \\ |\text{TP}| }$ & $\substack{ \Err_{\text{TP}}^{\text{o}} \\ \No_{\text{TP}}^{\text{o}} }$ & $\substack{ \Err_{\text{TP}}^{\text{v}} \\ \No_{\text{TP}}^{\text{v}} }$ & $\substack{ \Err_{\text{FN}} \\ |\text{FN}| }$ & $\substack{ \Err_{\text{FN}}^{\text{o}} \\ \No_{\text{FN}}^{\text{o}} }$ & $\substack{ \Err_{\text{FN}}^{\text{v}} \\ \No_{\text{FN}}^{\text{v}} }$ & $\substack{ \Err_{\text{FP}} \\ |\text{FP}| }$ & $\substack{ \Err_{\text{Sw}} \\ \text{Sw} }$ & fpps \\
    \midrule
MBM-\ourName & 0.287 & 0.298 & 0.348 & \bf 39.96 & $\substack{700.71 \\ 7320}$ &  $\substack{512.8 \\ 2007.3}$ &  $\substack{187.9 \\ 5312.7}$ & $\substack{5630.5 \\ 11261}$ & $\substack{4531.1 \\ 9062.2}$ & $\substack{1099.4 \\ 2198.8}$ & $\substack{457 \\ 914}$ & $\substack{440 \\ 44}$ & 1.57 \\[0.15cm]
  MBM-\solVo & 0.265 & 0.313 & 0.362 & 40.58 & $\substack{466.2 \\ 6119}$ & $\substack{369.6 \\ 1437.4}$ & $\substack{96.6 \\ 4681.6}$ & $\substack{6231 \\ 12462}$ & $\substack{4816.1 \\ 9632.2}$ & $\substack{1414.9 \\ 2829.8}$ & $\substack{517.5 \\ 1035}$ & $\substack{285 \\ 28.5}$ & 2.41 \\[0.15cm]
         MBM & 0.279 & 0.316 & 0.347 & 40.09 & $\substack{291.6 \\ 5751}$ &  $\substack{158.9 \\ 1007.1}$ &  $\substack{132.7 \\ 4743.9}$ & $\substack{6415 \\ 12830}$ & $\substack{5031.2 \\ 10062}$ & $\substack{1383.8 \\ 2767.5}$ & $\substack{246.5 \\ 493}$ & $\substack{330 \\ 33}$ & 2.41 \\[0.15cm]
      SORT & 0.320 & 0.313 & 0.364 & 40.15 & $\substack{315.9 \\ 6521}$ & $\substack{159.6 \\ 834.4}$ & $\substack{156.2 \\ 5686.6}$ & $\substack{6030 \\ 12060}$ & $\substack{5117.6 \\ 10235}$ & $\substack{912.4 \\ 1824.8}$ & $\substack{538 \\ 1076}$ & $\substack{425 \\ 42.5}$ & -- \\[0.05cm]
    \bottomrule
    &\\[-0.25cm]
        FastTracker & 0.827 & 0.581 & 0.679 & 24.77 & $\substack{620.2 \\ 17175}$ & $\substack{539.7 \\ 9778.7}$ & $\substack{80.5 \\ 7396.3}$ & $\substack{703.0 \\ 1406}$ & $\substack{645.4 \\ 1290.8}$ & $\substack{57.6 \\ 115.2}$ & $\substack{265.5 \\ 531}$ & $\substack{695.0 \\ 69.5}$ & -- \\
    \end{tabular}
\end{table*}

\begin{table*}
    \centering
    \caption{Results for MOT17-04 Video}
    \vspace{-0.1cm}
    \label{tab:MOT17-04}
    \begin{tabular}{cccc|c|ccc|ccc|c|c|c}
    \toprule
            & \multicolumn{3}{c|}{CV scores ($\uparrow$)} & \multicolumn{9}{c|}{TGOSPA metric ($\downarrow$) and its decomposition} & Hz ($\uparrow$) \\[0.1cm]
            & MOTA & HOTA & IDF1 & $d_{p}^{(c,\gamma)}$ & $\substack{ \Err_{\text{TP}} \\ |\text{TP}| }$ & $\substack{ \Err_{\text{TP}}^{\text{o}} \\ \No_{\text{TP}}^{\text{o}} }$ & $\substack{ \Err_{\text{TP}}^{\text{v}} \\ \No_{\text{TP}}^{\text{v}} }$ & $\substack{ \Err_{\text{FN}} \\ |\text{FN}| }$ & $\substack{ \Err_{\text{FN}}^{\text{o}} \\ \No_{\text{FN}}^{\text{o}} }$ & $\substack{ \Err_{\text{FN}}^{\text{v}} \\ \No_{\text{FN}}^{\text{v}} }$ & $\substack{ \Err_{\text{FP}} \\ |\text{FP}| }$ & $\substack{ \Err_{\text{Sw}} \\ \text{Sw} }$ & fpps \\
    \midrule
MBM-\ourName & 0.536 & 0.486 & 0.561 & \bf 50.31 & $\substack{836.0 \\ 26803}$ & $\substack{416.9 \\ 4137.1}$ & $\substack{419.1 \\ 22665.9}$ & $\substack{10377.0 \\ 20754}$ & $\substack{7022.0 \\ 14044.1}$ & $\substack{3355.0 \\ 6709.9}$ & $\substack{784.5 \\ 1569}$ & $\substack{590.0 \\ 59.0}$ & 1.33 \\[0.15cm]
  MBM-\solVo & 0.514 & 0.479 & 0.556 & 51.20 & $\substack{446.2 \\ 25397}$ & $\substack{165.4 \\ 3541.2}$ & $\substack{280.8 \\ 21855.8}$ & $\substack{11080.0 \\ 22160}$ & $\substack{7320.0 \\ 14640.0}$ & $\substack{3760.0 \\ 7520.0}$ & $\substack{1007.5 \\ 2015}$ & $\substack{600.0 \\ 60.0}$ & 1.35 \\[0.15cm]
         MBM & 0.506 & 0.466 & 0.536 & 51.46 & $\substack{863.4 \\ 25399}$ & $\substack{406.6 \\ 3801.2}$ & $\substack{456.8 \\ 21597.8}$ & $\substack{11079.0 \\ 22158}$ & $\substack{7190.0 \\ 14380.0}$ & $\substack{3889.0 \\ 7778.0}$ & $\substack{699.0 \\ 1398}$ & $\substack{655.0 \\ 65.5}$ & 1.66 \\[0.15cm]
      SORT & 0.540 & 0.498 & 0.564 & 50.39 & $\substack{290.9 \\ 26084}$ & $\substack{102.9 \\ 3409.4}$ & $\substack{187.9 \\ 22674.6}$ & $\substack{10736.5 \\ 21473}$ & $\substack{7385.9 \\ 14771.8}$ & $\substack{3350.6 \\ 6701.2}$ & $\substack{913.5 \\ 1827}$ & $\substack{695 \\ 69.5}$ & -- \\[0.05cm]
    \bottomrule
    &\\[-0.25cm]
        FastTracker & 0.988 & 0.887 & 0.968 & 14.19 & $\substack{209.8 \\ 47273}$ & $\substack{108.7 \\ 17939.7}$ & $\substack{101.1 \\ 29333.3}$ & $\substack{142.0 \\ 284}$ & $\substack{120.7 \\ 241.5}$ & $\substack{21.3 \\ 42.5}$ & $\substack{150.0 \\ 300}$ & $\substack{95 \\ 9.5}$ & -- \\
    \end{tabular}
\end{table*}

\begin{table*}
    \centering
    \caption{Results for MOT17-09 Video}
    \vspace{-0.1cm}
    \label{tab:MOT17-09}
    \begin{tabular}{cccc|c|ccc|ccc|c|c|c}
    \toprule
            & \multicolumn{3}{c|}{CV scores ($\uparrow$)} & \multicolumn{9}{c|}{TGOSPA metric ($\downarrow$) and its decomposition} & Hz ($\uparrow$) \\[0.1cm]
            & MOTA & HOTA & IDF1 & $d_{p}^{(c,\gamma)}$ & $\substack{ \Err_{\text{TP}} \\ |\text{TP}| }$ & $\substack{ \Err_{\text{TP}}^{\text{o}} \\ \No_{\text{TP}}^{\text{o}} }$ & $\substack{ \Err_{\text{TP}}^{\text{v}} \\ \No_{\text{TP}}^{\text{v}} }$ & $\substack{ \Err_{\text{FN}} \\ |\text{FN}| }$ & $\substack{ \Err_{\text{FN}}^{\text{o}} \\ \No_{\text{FN}}^{\text{o}} }$ & $\substack{ \Err_{\text{FN}}^{\text{v}} \\ \No_{\text{FN}}^{\text{v}} }$ & $\substack{ \Err_{\text{FP}} \\ |\text{FP}| }$ & $\substack{ \Err_{\text{Sw}} \\ \text{Sw} }$ & fpps \\
    \midrule
MBM-\ourName & 0.574 & 0.465 & 0.542 & \bf 19.35 & $\substack{138.2 \\ 3503}$ & $\substack{103.2 \\ 723.1}$ & $\substack{35.0 \\ 2779.9}$ & $\substack{911.0 \\ 1822}$ & $\substack{741.7 \\ 1483.4}$ & $\substack{169.3 \\ 338.6}$ & $\substack{34.5 \\ 69}$ & $\substack{175 \\ 17.5}$ & 2.21 \\[0.15cm]
  MBM-\solVo & 0.536 & 0.505 & 0.596 &19.71 & $\substack{140.6 \\ 3318}$ & $\substack{109.2 \\ 646.6}$ & $\substack{31.3 \\ 2671.4}$ & $\substack{1003.5 \\ 2007}$ & $\substack{779.9 \\ 1559.9}$ & $\substack{223.6 \\ 447.1}$ & $\substack{33.0 \\ 66}$ & $\substack{140 \\ 14}$ & 4.12 \\[0.15cm]
         MBM & 0.559 & 0.445 & 0.515 & 20.00 & $\substack{47.5 \\ 3055}$ & $\substack{25.1 \\ 388.0}$ & $\substack{22.5 \\ 2667.0}$ & $\substack{1135.0 \\ 2270}$ & $\substack{909.2 \\ 1818.5}$ & $\substack{225.8 \\ 451.5}$ & $\substack{21.5 \\ 43}$ & $\substack{160 \\ 16}$ & 3.43 \\[0.15cm]
      SORT & 0.530 & 0.437 & 0.501 & 20.69 & $\substack{25.4 \\ 2840}$ & $\substack{7.9 \\ 221.1}$ & $\substack{17.5 \\ 2618.9}$ & $\substack{1242.5 \\ 2485}$ & $\substack{992.7 \\ 1985.4}$ & $\substack{249.8 \\ 499.6}$ & $\substack{22.0 \\ 44}$ & $\substack{190 \\ 19}$ & -- \\[0.05cm]
    \bottomrule
    &\\[-0.25cm]
        FastTracker & 0.915 & 0.614 & 0.690 & 12.78 & $\substack{143.4 \\ 5173}$ & $\substack{105.4 \\ 2082.6}$ & $\substack{38.1 \\ 3090.4}$ & $\substack{76.0 \\ 152}$ & $\substack{61.9 \\ 123.9}$ & $\substack{14.1 \\ 28.1}$ & $\substack{49.5 \\ 99}$ & $\substack{195 \\ 19.5}$ & -- \\
    \end{tabular}
\end{table*}

As expected, FastTracker performs best in all the considered metrics and videos.
The algorithm ranking based on CV scores is difficult to justify clearly and is included for completeness. 
In the TGOSPA metric, the performance of the filters
is comparable to that of SORT, among which the MBM-\ourName{} is the best.
Detailed observations and discussions follow.

\begin{observation}%
[More informative \probabilityOfDetection{} model enhances measurement-to-track associations]\label{observation:more-TPs}
    Both MBM-\ourName{} and MBM-\solVo{} mostly have a higher number $|\text{TP}|$ compared to the basic MBM in all videos.
    When pedestrians are visible, the occlusion-handling strategies assign considerably higher \expectedProbabilityOfDetection{} to marks compared to $P_D {=} 0.529$ used otherwise.
    Measurement\discretionary{-}{-}{-}to\discretionary{-}{-}{-}track associations are thus more likely for the MBM\discretionary{-}{-}{-}\ourName{} and MBM\discretionary{-}{-}{-}\solVo{} compared to the basic MBM.
\end{observation}
    
\begin{observation}%
[\ourName{} and \solVo{} computational complexities are hardly predictable]
    The algorithms' computational complexity was measured in frames processed per second (fpps), which is denoted with \emph{Hz} in the MOT-17 website\footnote{
        The values for SORT and FastTracker of $143.3$ Hz and $355.1$ Hz, respectively, are reported by the website \url{motchallenge.net/results/MOT17}.
        The values were provided by the algorithms' authors for the entire dataset and not officially evaluated by the MOTChallenge.
    }.
    Since \probabilityOfDetection{} for visible pedestrians are considerably higher when using either occlusion-handling strategy, the discrepancy among different measurement-to-track associations is larger.
    This may lead to better computational efficiency for a small number of objects, such as for MBM-\solVo{} in MOT17-09.
    For a larger number of objects, however, the computational demand is likely to be larger when using either \ourName{} or \solVo{}.
    It should be noted that implementations of MBM-\ourName{}, MBM-\solVo{}, and the MBM filters was run in R2025b version of Matlab\textregistered{} software on an Apple 
    M4 Air laptop. 

    Median times spent on a single call of \solVo{} and \ourName{} 
    blocks of code,
    i.e., median times of the \probabilityOfDetection{} computation for all marks for a single time step, are given in Table~\ref{tab:PoDcomptTimes}. 
    As expected, the MBM-\ourName{} computes all $P_D(m)$ values (for a frame) slower than the MBM-\solVo{} in all videos.
\end{observation}

\begin{table}[h]
    \centering
    \vspace{-0.3cm}
    \caption{
        Complexities of \solVo{} and \ourName{} per frame. 
    }\vspace{-0.1cm}
    \label{tab:PoDcomptTimes}
    \begin{tabular}{cccc}
         & MOT17-02 & MOT17-04 & MOT17-09 \\\hline
         median time \ourName{} [ms] & 225 & 304 & 186 \\
         median time \solVo{} [ms] & 56 & 182 & 36
    \end{tabular}
    \vspace{-0.3cm}
\end{table}

\begin{observation}%
    [\ourName{} increases tracking performance for visible pedestrians over \solVo{}]
    The MBM-\ourName{} tracks a larger number $\No_{\text{TP}}^{\text{v}}$ of \emph{visible} pedestrians with a similar error $\Err_{\text{TP}}^{\text{v}}$ as MBM-\solVo{} and MBM, and also as SORT except for MOT17-02 video. 
\end{observation}

\begin{observation}%
    [\solVo{} preserves tracking performance for visible pedestrians]
    The MBM-\solVo{} tracks a similar or smaller number $\No_{\text{TP}}^{\text{v}}$ of \emph{visible} pedestrians with a similar or smaller error $\Err_{\text{TP}}^{\text{v}}$ compared to MBM\discretionary{-}{-}{-}\ourName{} and MBM.
    The latter indicates that MBM\discretionary{-}{-}{-}\solVo{} provides less outliers. 
\end{observation}

\begin{observation}[\ourName{} handles more occlusions than \solVo{}]\label{observation:POR-handles-more-occluded-pedestrians}
    Among the properly tracked objects, MBM\discretionary{-}{-}{-}\ourName{} tracks more occluded pedestrians than MBM\discretionary{-}{-}{-}\solVo{} in all videos. 
    This follows since the ratios $\sfrac{ \No_{\text{TP}}^{\text{o}} }{ |\text{TP}| }$ are higher for MBM-\ourName{} compared to MBM-\solVo{}:
    in \mbox{MOT17-02} we have $0.274 {>} 0.235$,
    in \mbox{MOT17-04} it is $0.154 {>} 0.139$, and
    in \mbox{MOT17-09} it is $0.206 {>} 0.195$.
\end{observation}

\begin{observation}[\ourName{} misses less pedestrians]
    MBM\discretionary{-}{-}{-}\ourName{} has less $|\text{FN}|$, $\No_{\text{FN}}^{\text{o}}$ and also $\No_{\text{FN}}^{\text{v}}$ than MBM\discretionary{-}{-}{-}\solVo{} and MBM in all the videos.
    That is, MBM\discretionary{-}{-}{-}\ourName{} misses fewer pedestrians under any of the considered circumstances compared to the other filters.
    Furthermore, MBM-\ourName{} mitigates $\No_{\text{FN}}^{\text{v}}$ over SORT in MOT17-09, has nearly the same $\No_{\text{FN}}^{\text{v}}$ as SORT in MOT17-04, however, it has larger $\No_{\text{FN}}^{\text{v}}$ than SORT in MOT17-02.
    Nevertheless, MBM-\ourName{} misses less occluded pedestrians then SORT as it has fewer $\No_{\text{FN}}^{\text{o}}$, and also in total $|\text{FN}|$.
\end{observation}

\begin{observation}%
    [\solVo{} may fail handling occlusions]
    In MOT17-04, MBM-\solVo{} provides the least number $\No_{\text{TP}}^{\text{o}}$ of occluded pedestrians.
    At the same time, it has the largest $|\text{FP}|$.
    That is, the use of the \solVo{} strategy to handle occlusions may lead to reduced performance.
\end{observation}

\begin{observation}[More estimates leads to more false positives]
    MBM-\ourName{} and MBM-\solVo{} have more overall number $|\text{TP}| {+} |\text{FP}|$ of estimates than MBM.
    While both MBM-\ourName{} and MBM-\solVo{} usually have more TPs (Observation~\ref{observation:more-TPs}), they also have more FPs relative to MBM.
    In either \ourName{} or \solVo{} strategy, high existence probabilities of occluded Bernoullies last longer and yield distant predictive estimates.
    Whenever such estimates are not assigned to \groundTruth{}s in TGOSPA minimization, they get counted as FPs.
    This issue is present in FastTracker as well, especially in MOT17\discretionary{-}{-}{-}04, where the error $\Err_{\text{FP}} {=} 150$ is only slightly lower than $\Err_{\text{TP}} {=} 209.8$, and it has higher influence on the metric value compared to $\Err_{\text{FN}}$ and $\Err_{\text{Sw}}$.
\end{observation}

\begin{observation}[Handling occlusions $\neq$ handling switches]
    Although a switch may have several graphical explanations~\cite{KrKoStXiSvFe:TGOSPA-params-visual:ArXiv}, at least one $\text{Sw}$ appears whenever a \groundTruth{} pedestrian is successively successfully tracked by two different trajectories (their fragments) for at least $10$ frames by each trajectory.
    It turns out that the algorithms get rewarded more for successfully estimating occluded objects than they get penalized for an occasional switch. 
    Note that if $\gamma$ was set even larger, such track fragments may fail to be counted as $\text{Sw}$~\cite{KrKoStXiSvFe:TGOSPA-params-visual:ArXiv}.
    Although FastTracker handles occlusions, it has the largest number of switches in MOT17-02 (and also in MOT17-09).
    Furthermore, the largest TGOSPA error term for FastTracker in MOT17-09 is the $\Err_{\text{Sw}}$.
    Except for MOT17-04, MBM-\ourName{} has more switches than MBM-\solVo{}.
    Nevertheless, the filters mostly have fewer switches than SORT.
\end{observation}

\begin{observation}%
[Model-based approach can beat ad-hoc-based methods]
    Utilizing the same input data, the MBM-\ourName{} beats SORT in all the videos by improving the performance of the basic MBM.
    That is, the use of a more realistic model led to better results.
    However, algorithms such as FastTracker will be hard to outperform since they process considerably larger amounts of information (entire images).
    It should be noted that such ad-hoc-based algorithms are usually unable to reliably quantify their estimation errors, e.g., in probabilistic terms, since they do not utilize probabilistic models which may be problematic for safety-critical applications.
\end{observation}\vspace{0.1cm}

The experiment indicated that the proposed occlusion\discretionary{-}{-}{-}handling strategy \ourName{} improves tracking performance not only under occlusion, but also for visible pedestrians, and that it also mitigates false negatives.
The strategy, however, provides more false positives and switches.
On the other hand, this effect also appears in advanced ad-hoc-based algorithms processing not only BB detections such as the FastTracker. 
The proposed strategy thus appears promising for dealing with occlusions in visual tracking. 

\section{Applicability Among Different Sensors} 
While the considered framework applies among many sensors~\cite{Mahler-Book:2014}, the SPO\discretionary{-}{-}{-}D model utilized in this work requires the sensed data to be segmented into \emph{clusters}, c.f.,~\cite{Wyffels:NegativeOcclu:2015,Dai-Occlusion:2020}.
Furthermore, at most one cluster per object must exist in the set of clusters for each time step.
This section elaborates on the generality of this approach for dealing with occlusions using various sensing equipment.
Note that obtaining reliable model parameters is essential to enable \emph{fully model-based} tracking. 

\subsection{Camera-Based Tracking}
In the visual tracking example considered before, the said clusters are BBs given by the RFCNN detector.
Many alternative BB detectors exist, 
e.g.,~\cite{MOT-16:2016}. 
To find suitable measurement model parameters \emph{for the chosen detector}, including the \probabilityOfDetection{} as a function of the visibility, 
the method from~\cite{KrKoSt:2023_FUSION} can be utilized.
Note that objects such as vehicles can be tracked visually as well, for which suitable motion and measurement models must be given.
Furthermore, object class likelihoods (if provided by the detector) can be incorporated into the tracking algorithm~\cite{Krejci:Feature-based:2022}.

With pixel-wise segmentation, e.g.,~\cite{Wu:OccluParts:2006},
clusters take the form of sets of pixels covering a potential object.
Although the object visibility becomes physically straightforward,
the \probabilityOfDetection{} function model must be developed. 

Stereo cameras further output depth measurements,
e.g.,~\cite{StereoCam3DTracking:2009}.
Such measurements can be easily accounted for in the considered tracking framework by adding a suitable equation alongside the BB measurement model. 
Direct depth measurements naturally result in considerably higher depth estimation accuracy and thus better visibility computation accuracy, c.f., Eq.~\eqref{eq:visibility-ratio}.\\

Employing sensors other than video cameras can apparently be done along the lines of the above text.
\subsection{Laser-Based Tracking}

LiDAR point clouds
contain range measurements, e.g.,~\cite{Chen:LidarOccluMap:2018}.
Employing suitable segmentation algorithm, the resulting clusters can be treated analogously to stereo cameras.
Furthermore, Doppler measurements may be available depending on the considered measurement unit.
The \probabilityOfDetection{} function model would depend on the chosen segmentation algorithm. 

Laser range finder could be understood as a special case of LiDAR that lacks the vertical axis, e.g.,~\cite{Occlusions-MHT:2014,Dai-Occlusion:2020}. 
Since~\cite{Dai-Occlusion:2020} utilized the GLMB tracking framework, most parts of the model developed therein can be adopted directly for the proposed MBM-\ourName{} filter. 




%
%
\section{Conclusion}\label{sec:Conclusion}
This paper focused on multi-object tracking problems
where occlusions among objects may arise. A principled approximation was proposed, leading to a straightforward and fully model-based strategy for handling occlusions.
The essence of the strategy is the calculation of the conditional expectation related to detection probability. As the solution necessitates marked state densities to be proven optimal, the paper investigated the implementation of filters using multi-Bernoulli mixtures with distinct marks.
The application of visual tracking demonstrated that the proposed strategy outperformed others.
Future work could focus on incorporating Poisson birth or addressing implementation for unmarked objects.


\appendices
\section{Rigorous Palm Conditioning}\label{app:Palm-definition}
A point process could be understood as a \emph{collection of points drawn at random}.
When the number of points is locally finite and they do not coincide, the point process can be represented as an RFS.
However, counting measure representation is most common in the literature discussing Palm conditioning, see~\cite[Ch.~13, Ch.~15.5]{Vere-Jones:2008}, \cite[Sec.~3.2]{StochasticGeometry-lecturesBook:2004} for details.
In this appendix, the counting measure treatment is employed to incorporate Palm conditioning into FISST, i.e., the theory used in this paper.

\subsection{Preliminaries}
In general~\cite{Mahler-Book:2014}, the underlying spaces $\X$ and $\Z$ can be assumed \emph{locally compact}, \emph{second countable}, and \emph{Hausdorff} (\locallyCompactSecondCountableHausdorff{}) topological spaces.
With such an assumption, one can rigorously tackle applications involving more general spaces than $\mathbb{R}^n$~\cite[Chapter~18]{Mahler-Book:2014}.
For instance, objects may move on a manifold, such as on a globe, or objects that live in state spaces of different dimensions may coexist.
\locallyCompactSecondCountableHausdorff{} spaces are Polish~\cite[p.~13~and~29]{Kechris:DescriptiveSetTheory:1995}, i.e., any \locallyCompactSecondCountableHausdorff{} space can be turned into a \emph{complete separable metric} (\completelySeparableMetric{}) space.
A metric on the underlying state space is usually needed to measure the performance of tracking algorithms\footnote{
    The metric must be chosen such that the space is \completelySeparableMetric{} in that metric.
    Being Polish, such a metric is guaranteed to exist for a \locallyCompactSecondCountableHausdorff{} space.
    If $\X {=} \mathbb{R}^n$, any metric on $\mathbb{R}^n$, including the cut-off metric, can be used.
}~\cite{TrajectoryGOSPA:2020}.
Therefore, being \completelySeparableMetric{} seems as the convenient assumption for $\X$ and $\Z$.

The \emph{Dirac measure centered at} $\mathbf{x} {\in} \X$, denoted with $\delta_{\mathbf{x}}$ is the set function, such that $\delta_{\mathbf{x}}(B) {=} 1$ if $\mathbf{x} {\in} B {\subset} \X$ and zero otherwise.
The \emph{counting measure} $N_{\Xi}$ \emph{associated with} $\Xi {\subset} \X$ is a set function that simply counts the number of points of $\Xi$ within $B$, i.e.,
\begin{align}
    N_{\Xi}(B) = \textstyle \sum_{\mathbf{x}\in\X} \delta_{\mathbf{x}}(B) = |\Xi \cap B| \, .
    \label{eq:counting-measure-def}
\end{align}

Let $(\mathfrak{X}, \mathfrak{S}, P_{\Xi})$ be the probability space corresponding to the RFS $\Xi$.
Briefly, the set $\mathfrak{X} {=} \uplus_{n\geq0} \X^{(n)}$ is the space such that $\Xi {\in} \mathfrak{X}$, where $B^{(n)}$ is the set of all subsets of the closed set $B {\subseteq} \X$ containing $n$ elements and $B^{(0)} {=} \{ \emptyset \}$ is the empty configuration.
The set $\mathfrak{S}$ is a convenient sigma-algebra on $\mathfrak{X}$ (see~\cite[Appendix~F]{Mahler-Book:2007}, \cite[Ch.~4]{Mahler:1997}) and $P_{\Xi}$ is the probability measure.

To give the RPD, the following definitions are needed.
The \emph{first moment measure} $D_{\Xi}$ is defined as 
\begin{align}
    D_{\Xi}(B) = \Expect{ N_{\Xi}(B) } \,,
    \label{app:first-moment-measure}
\end{align}
where $\Expect{\cdot}$ is the expectation operator.
The first moment measure of any RFS must be absolutely continuous with respect to (w.r.t.)~some reference measure, otherwise it fails to form a locally finite set~\cite[p.~138]{Vere-Jones:2003}, thus
\begin{align}
    D_{\Xi}(B) = \textstyle \int_B D_{\Xi}(\mathbf{x}) \, \d\mathbf{x}\,,
    \label{app:PHD-absolutely-continuous}
\end{align}
where $D_{\Xi}(\mathbf{x})$ is the PHD~\eqref{eq:PHD-definition} of the RFS $\Xi$. 
The same holds true for the so-called \emph{Janossy densities}~\cite[p.136]{Vere-Jones:2003}, which are in the FISST literature defined via \emph{set derivatives} $\tfrac{\delta}{\delta X}$~\cite[Ch.~4]{Mahler:1997} as
\begin{align}
    p_{\Xi}(X) &= \textstyle \frac{\delta \beta_{\Xi} (\,\cdot\,)}{\delta X}, &
    \beta_{\Xi}(G) &= \textstyle P_{\Xi} \big( \uplus_{n\geq0} G^{(n)} \big),
\end{align}
where $\beta_{\Xi}(G)$ is the \emph{belief mass function} evaluated at the closed set $G {\subseteq}\X$ and $p_{\Xi}(\{\mathbf{x}^1,\dots,\mathbf{x}^n\})$ is the $n$-th order Janossy density aka the FISST density function.
Probabilities thus can be computed as~\cite[p.~714]{Mahler-Book:2007} 
\begin{align}
    P_{\Xi}(S) = \textstyle \int_{ \chi^{-1} ( S \cap \X^{(|X|)} ) } p_{\Xi}(X) \delta X \, ,
\end{align}
where $\chi$ is the mapping from vectors to sets defined as $\chi( [\mathbf{x}^1,\dots,\mathbf{x}^n]\T ) = \{\mathbf{x}^1,\dots,\mathbf{x}^n\}$.

The \emph{Campbell measure} $C_{\Xi}$ corresponding to $\Xi$ is the measure on $\X \times \mathfrak{X}$ satisfying (cf.~\cite[pp.~270-271]{Vere-Jones:2008})
\begin{subequations}
\begin{align}
    C_{\Xi}(B \times S) &= \Expect{ N_{\Xi}(B) \cdot \mathbf{1}_{S} } \\
    &= \textstyle \int_{ \chi^{-1} ( S \cap \X^{(|X|)} ) } |X\cap B| \, p_{\Xi}(X) \delta X \, , \label{app:Campbell-set-integral}
\end{align}
\end{subequations}
for each measurable $B {\subseteq} \X$ and each event $S {\in} \mathfrak{S}$, where $\mathbf{1}_{S}$ is the indicator random variable of the event $S {\in} \mathfrak{S}$.

\subsection{Palm Conditioning within FISST}
The \emph{Palm distribution} $P_{\Xi | \mathbf{x}\in\Xi }$ is defined as the Radon\discretionary{-}{--}{--}Nikodým derivative of the Campbell measure in the first argument w.r.t~the first moment measure $D_{\Xi}$~\eqref{app:first-moment-measure}.
Since $D_{\Xi}$~\eqref{app:first-moment-measure} is absolutely continuous, it follows that for any fixed event $S {\in} \mathfrak{S}$, 
\begin{align}
    C_{\Xi}( B \times S ) 
    &= \textstyle \int_{B} P_{\Xi | \mathbf{x}\in\Xi }( S | \mathbf{x} ) D_{\Xi}(\mathbf{x}) \,\d\mathbf{x}
    \,,
\end{align}
which can be understood as the \emph{disintegration} of $C_{\Xi}$ w.r.t.~the first component. 
Note that $P_{\Xi | \mathbf{x}\in\Xi }( S | \mathbf{x} )$ is a probability measure in $S$ and a function in $\mathbf{x}$.
Its density can be defined using several methods.
In the point process literature, it is common to take the Radon\discretionary{-}{--}{--}Nikodým derivative of $P_{\Xi | \mathbf{x}\in\Xi }( \,\cdot\, | \mathbf{x} )$ w.r.t.~the \emph{unit-rate Poisson point process}~\cite[Ch.~10.4]{Vere-Jones:2008}. 
In FISST, however, belief mass functions are used instead. 

For a fixed measurable $B {\subseteq} \X$ define 
\begin{subequations}
\begin{align}
    \beta_{C,\Xi}( G ; B ) &\triangleq C_{\Xi}\big( B \times ( \cup_{n\geq0} G^{(n)} ) \big) \label{app:belief-mass-from-Campbell} \\
    & \hspace{-0.5cm} = \textstyle \int_{B} \underbrace{ P_{\Xi | \mathbf{x}\in\Xi }\big( ( \cup_{n\geq0} G^{(n)} ) | \mathbf{x} \big) }_{ \triangleq \beta_{ \Xi | \mathbf{x\in\Xi} } (G | \mathbf{x}) } D_{\Xi}(\mathbf{x}) \,\d\mathbf{x} \,,
\end{align}
\end{subequations}
where $\beta_{C,\Xi}$ is the belief mass function corresponding to $C_{\Xi}$ in the second argument and $\beta_{ \Xi | \mathbf{x\in\Xi} }$ is the belief mass function corresponding to $P_{\Xi | \mathbf{x}\in\Xi }$ with $\mathbf{x}$ fixed.
Note that $\beta_{C,\Xi}( G ; B )$~\eqref{app:belief-mass-from-Campbell} can be used instead of $C_{\Xi}( B \times S )$ thanks to the Choquet theorem, see~\cite[p.~713]{Mahler-Book:2007}, \cite[Ch.3]{Mahler:1997}.
Assuming it exists, the set derivative of $\beta_{C,\Xi}( G ; B )$~\eqref{app:belief-mass-from-Campbell} where $B$ is still fixed is denoted with
\begin{align}
    p_{C,\Xi}(X; B) &\triangleq \textstyle \frac{ \delta \beta_{C,\Xi}( \,\cdot\, ; B ) }{ \delta X } 
    = \int_{B} \underbrace{ \textstyle \frac{ \delta \beta_{\Xi | \mathbf{x}\in\Xi } ( \,\cdot\, | \mathbf{x} ) }{ \delta X } }_{ \triangleq p_{\Xi | \mathbf{x}\in\Xi } ( X | \mathbf{x} ) } \, D_{\Xi}(\mathbf{x}) \,\d\mathbf{x} \, , \notag\\[-0.6cm]
    \label{app:set-derivative-of-beta}
\end{align}
where $X {\subset} \X$ is a finite set such that $\mathbf{x} {\in} X$ and $\mathbf{x} {\in} B$, while $p_{\Xi | \mathbf{x}\in\Xi }$ is the FISST density corresponding to $\beta_{ \Xi | \mathbf{x\in\Xi} }$, i.e., the Palm density.
Substituting $X {=} O {\cup} \{\mathbf{x}\}$ into~\eqref{app:set-derivative-of-beta} yields
\begin{align}
    p_{C,\Xi}( O {\cup} \{\mathbf{x}\} ; B) &= \textstyle \int_{B}  \underbrace{ p_{\Xi | \mathbf{x}\in\Xi } ( O {\cup} \{\mathbf{x}\} | \mathbf{x} ) }_{ \triangleq p_{ \Theta | \mathbf{x} \in \Xi } ( O | \mathbf{x} ) }  \, D_{\Xi}(\mathbf{x}) \d\mathbf{x} ,
    \notag\\[-0.6cm]
    \label{app:PC-Xi-B}
\end{align}
where the function $O \mapsto p_{\Xi | \mathbf{x}\in\Xi } ( O {\cup} \{\mathbf{x}\} | \mathbf{x} )$ for a given $\mathbf{x}$ is defined as the desired \emph{reduced Palm density} (RPD), i.e.,
\begin{align}
    p_{ \Theta | \mathbf{x} \in \Xi } ( O | \mathbf{x} ) &\triangleq p_{\Xi | \mathbf{x}\in\Xi } ( O {\cup} \{\mathbf{x}\} | \mathbf{x} ) \, ,
\end{align}
with $\Theta {=} \Xi {\setminus} \{\mathbf{x}\}$ being the RFS of points in $\Xi$ besides $\mathbf{x}$.

The relation~\eqref{eq:RPD-RFS-definition} suitable for practical computation of RPD remains to be established.
Employing~\eqref{app:Campbell-set-integral} directly, an alternative form of~\eqref{app:belief-mass-from-Campbell} involves the FISST density,
\begin{subequations}
\begin{align}
    \beta_{C,\Xi}(G; B) 
    &= \textstyle \int_G \big( \sum_{\mathbf{x} \in X} \delta_{\mathbf{x}}(B) \big) \, p_{\Xi}(X) \, \delta X \\
    &= \textstyle \int_B \int_G p_{\Xi}(X \cup \{\mathbf{x}\}) \, \delta X \, \d \mathbf{x},
    \label{app:belief-mass-from-Campbell-alternative}
\end{align}
\end{subequations}
where the representation~\eqref{eq:counting-measure-def} was used and the last two equations follow as a simple generalization of the proof of~\cite[Theorem~2]{Mahler-PHD:2003}, i.e., a straightforward extension of~\cite[Eq.~(4.74)]{Mahler-Book:2007} from Dirac ``functions'' to measures.

Using the fundamental theorem of multi-object calculus~\cite[pp.~159-161]{Mahler:1997}, the set derivative~\eqref{app:set-derivative-of-beta} of $\beta_{C,\Xi}(G; B)$~\eqref{app:belief-mass-from-Campbell-alternative} thus also equals to
\begin{align}
    p_{C,\Xi}(X; B) &= \textstyle \frac{ \delta \beta_{C,\Xi}( \,\cdot\, ; B ) }{ \delta X } = \int_B \, p_{\Xi}(X \cup \{\mathbf{x}\}) \d\mathbf{x} \, ,
    \label{app:alternative-PC-Xi-B}
\end{align}
provided that $p_{\Xi}(X) {<} M$ for some large $M {\in} \mathbb{R}^+$ for all $X {\in} \mathfrak{X}$ using the Lebesgue dominated convergence theorem.
Substituting $X {=} O {\cup} \{\mathbf{x}\}$ into~\eqref{app:alternative-PC-Xi-B} and using~\eqref{app:PC-Xi-B} yields
\begin{align}
    \textstyle \int_B p_{\Xi}( O {\cup} \{\mathbf{x}\} ) \, \d\mathbf{x} = \int_{B}  p_{ \Theta | \mathbf{x} \in \Xi } ( O | \mathbf{x} ) \, D_{\Xi}(\mathbf{x}) \, \d\mathbf{x} \, ,
\end{align}
which must hold for any measurable $B {\subseteq} \X$ and thus the integrands must be equal almost everywhere (provided that $\X$ is \emph{directionally limited}~\cite[p.~7]{Heinonen:AnalysisMetricSpaces:2001} using the~Lebesgue differentiation theorem).
Assuming the densities are continuous, the integrands are equal everywhere in $\X$, thus 
\begin{align}
    p_{\Xi}( O {\cup} \{\mathbf{x}\} ) = p_{ \Theta | \mathbf{x} \in \Xi } ( O | \mathbf{x} ) \, D_{\Xi}(\mathbf{x}) \, ,
\end{align}
for any $\mathbf{x} {\in} \X$.
Dividing both sides by $D_{\Xi}(\mathbf{x})$, while assuming it is nonzero, yields the desired relation~\eqref{eq:RPD-RFS-definition}.

\section{Proof of Proposition~\ref{prop:best-fitting-KLD} and Its Special Case}\label{app:proof-of-proposition-KLD}
The proof of Proposition~\ref{prop:best-fitting-KLD} uses the following Lemma, c.f.~\cite[p.48]{StochasticGeometry-lecturesBook:2004}, \cite[p.~512]{Vere-Jones:2008}. 

\begin{lemma}[Campbell-Mecke Formula for RFSs]\label{lemma-Palm-expectation}
    Let a function $f: \X {\times} \mathfrak{X} {\rightarrow} \mathbb{R}^+$ be measurable w.r.t.~the Campbell measure.
    Let $\RFSXi$ be an RFS on $\X$ with a density $p_{\RFSXi}(X)$ and PHD $D_{\RFSXi}(\mathbf{x})$.
    Then 
    \begin{align}
        \!\! \textstyle \mathrm{E}_{\RFSXi} \! \big[ \! \sum_{\mathbf{x} \in \RFSXi} \! f(\mathbf{x}, \RFSXi {\setminus} \{\mathbf{x}\} ) \big] 
        {=}  \textstyle \int \! \mathrm{E}_{\RFSTheta | \mathbf{x}} \! \big[ f(\mathbf{x}, \RFSTheta) \big] \! D_{\RFSXi}(\mathbf{x}) \d \mathbf{x} . \!\! \label{app:eq:lemma-Palm-expectation}
    \end{align}
    \textsc{Proof:}
    Directly,
    \begin{subequations}
    \begin{align}
        &\!\!\!\! \textstyle \mathrm{E}_{\RFSXi} \! \big[ \! \sum_{\mathbf{x} \in \RFSXi} \! f(\mathbf{x}, \RFSXi {\setminus} \{\mathbf{x}\} ) \big] \notag 
        {=} \! \textstyle \int \! \sum_{\mathbf{x}\in X} \! f(\mathbf{x}, X {\setminus} \{\mathbf{x}\}) p_{\RFSXi}(X) \delta\! X \\
        & \hspace{0.2cm} = 
        \textstyle \sum_{n=0}^\infty \tfrac{1}{n!} \sum_{i=1}^n \int\int\cdots\int\int\cdots\int \notag \\
        & \hspace{1.4cm} f(\mathbf{x}^i, \{\mathbf{x}^1,\dots,\mathbf{x}^{i-1},\mathbf{x}^{i+1},\dots,\mathbf{x}^n\}) \times \notag \\
        & \hspace{1.4cm} p_{\RFSXi} (\{\mathbf{x}^1,\dots,\mathbf{x}^{i-1},\mathbf{x}^{i+1},\dots,\mathbf{x}^n\} \cup \{\mathbf{x}^i\}) \times \notag \\
        & \hspace{2.1cm} \d\mathbf{x}^1 \cdots \d\mathbf{x}^{i-1}\d\mathbf{x}^{i+1}\cdots\d\mathbf{x}^n \d\mathbf{x}^i \, .
    \end{align}
    \end{subequations}
    For each $n$, the integrals for $i {=} 1,\dots,n$ are the same. 
    Noting that the summand for $n{=}0$ can be neglected, 
    \begin{subequations}
    \begin{align}
        &\textstyle \mathrm{E}_{\RFSXi} \! \big[ \sum_{\mathbf{x} \in \RFSXi} \! f(\mathbf{x}, \RFSXi {\setminus} \{\mathbf{x}\} ) \big] \notag \\
        &\textstyle \hspace{0.2cm} = \int \sum_{n=1}^{+\infty} \tfrac{n}{n!} \int \cdots \int f(\mathbf{x}, \{\mathbf{o}^1,\dots,\mathbf{o}^{n-1}\}) \times \notag \\
        &\textstyle \hspace{0.7cm} p_{\RFSXi} ( \{\mathbf{o}^1,\dots,\mathbf{o}^{n-1}\} \cup \{\mathbf{x}\} ) \d\mathbf{o}^1\cdots\d\mathbf{o}^{n-1} \cdot \d\mathbf{x} \\
        &\textstyle \hspace{0.2cm} = \int \sum_{m=0}^{+\infty} \tfrac{1}{m!} \int \cdots \int f(\mathbf{x}, \{\mathbf{o}^1,\dots,\mathbf{o}^{m}\}) \times \notag \\
        &\textstyle \hspace{1.0cm} \underbrace{ p_{\RFSXi} ( \{\mathbf{o}^1,\dots,\mathbf{o}^{m}\} \cup \{\mathbf{x}\} ) }_{ p_{\RFSTheta|\mathbf{x}}(\{\mathbf{o}^1,\dots,\mathbf{o}^{m}\} | \mathbf{x}) D_{\RFSXi}(\mathbf{x}) } \d\mathbf{o}^1\cdots\d\mathbf{o}^{m} \cdot \d\mathbf{x} \\
        &\textstyle \hspace{0.2cm} = \int \int f(\mathbf{x}, O) p_{\RFSTheta|\mathbf{x}}(O|\mathbf{x}) \delta O \, D_{\RFSXi}(\mathbf{x})  \d \mathbf{x} \, ,
    \end{align}
    \end{subequations}
    using substitution $m {=} n {-} 1$ and the RPD~\eqref{eq:RPD-RFS-definition}. \hfill $\square$
\end{lemma}

\begin{corollary}\label{corollary-Campbell-Mecke-Marked}
    If the space $\X$ is a joint space with marks $\X {=} \X^\projFcn {\times} \spaceOfMarks$ and the RFS $\RFSXi$ is a marked RFS on $\X^\projFcn$ with marks in $\spaceOfMarks$, then~\eqref{app:eq:lemma-Palm-expectation} from Lemma~\ref{lemma-Palm-expectation} becomes
    \begin{align}
        & \textstyle \mathrm{E}_{\RFSXi} \big[ \sum_{\mathbf{x} \in \RFSXi} f(\mathbf{x}, \RFSXi {\setminus} \{\mathbf{x}\} ) \big]  \\
        &\hspace{0.5cm} 
        = \textstyle \sum_{m \in \spaceOfMarks} \mathrm{E}_{\RFSTheta,x|(\,\cdot\,,m)} \big[ f\big( (x,m), \RFSTheta \big) \big] \cdot D(m) \, , \notag
    \end{align}
    where the inner expectation is a function of $m$, computed over the density $p_{\RFSTheta | \mathbf{x}} \big( O| (x,m) \big) {\cdot} D(x|m)$ for each $m$ fixed, and the PHD of $\RFSXi$ is $D_{\RFSXi}(x,m)=D(x|m)\cdot D(m)$, see~\eqref{eq:PHD-decomposition-xk,m}.
\end{corollary}

\noindent
\textsc{Proof of Proposition~\ref{prop:best-fitting-KLD}:}
First, from the structure of $Q$~\eqref{eq:space-of-functions-with-parameter}, the optimal parameter can be established as
\begin{align}
    \paramPD^* &= \underset{ \paramPD \, : \, \X {\rightarrow} [0,1] }{ \arg\min } \, D_{KL} \big( p \, \| \, q_\paramPD \big) \, . \label{app:eq:KLD-mini-aux}
\end{align}
Given the definitions of $p(\,\cdot\,)$~\eqref{eq:true-joint} and $q(\,\cdot\,, \paramPD)$~\eqref{eq:joint-with-parameter}, notice that the KLD in~\eqref{app:eq:KLD-mini-aux} can be written as
\begin{align}
    \!\! D_{KL} \big( p \, \| \, q_\paramPD \big) {=}  \mathrm{E}_{X_k} \! \scalebox{1.3}{\Big[} \mathrm{E}_{Z_k|X_k} \! \Big[ \log \tfrac{p_{ \mathrm{A}\text{-}\mathrm{SPO}\text{-}\mathrm{D}_k } ( Z_k | X_k )}{p_{ \mathrm{A}\text{-}\mathrm{SPO}_k } ( Z_k | X_k,\, \paramPD )} \! \Big] \scalebox{1.3}{\Big]} \! . \!
\end{align}
The following adjustments of the inner expectation follow the classic proof~\cite[p.~277]{KollerFriedman-book:2009}.
We have that
\begin{align}
    \label{app:eq:inner-expectation-adjustment-one}
    &  \mathrm{E}_{Z_k|X_k} \! \Big[ \! \log \! \tfrac{p_{ \mathrm{A}\text{-}\mathrm{SPO}\text{-}\mathrm{D}_k } ( Z_k | X_k )}{p_{ \mathrm{A}\text{-}\mathrm{SPO}_k } ( Z_k | X_k,\, \paramPD )} \! \Big] = \\
    & \hspace{1.7cm} \mathrm{E}_{Z_k|X_k} \! \Big[ \! \log \! \tfrac{p_{ \mathrm{A}\text{-}\mathrm{SPO}\text{-}\mathrm{D}_k } ( Z_k | X_k )}{p_{ \mathrm{A}\text{-}\mathrm{SPO}_k } ( Z_k | X_k,\, \paramPD^{\text{g}} )} \! \Big] \notag\\
    & \hspace{1.7cm} + \mathrm{E}_{Z_k|X_k} \! \Big[ \! \log \! \tfrac{ p_{ \mathrm{A}\text{-}\mathrm{SPO}_k } ( Z_k | X_k,\, \paramPD^{\text{g}} ) }{ p_{ \mathrm{A}\text{-}\mathrm{SPO}_k } ( Z_k | X_k,\, \paramPD ) } \! \Big] , \notag
\end{align}
where the \emph{guessed} $\paramPD^g(\mathbf{x}_k) {=} \mathrm{E}_{O_k|\mathbf{x}_k} [ \, P_\ObjectGenRFS(\mathbf{x}_k, O_k) \, ]$ is to be proven to be the optimal $\paramPD^*$.
The first term on the right-hand-side of~\eqref{app:eq:inner-expectation-adjustment-one} does not depend on $\paramPD$ and since KLD is nonnegative, it can be neglected.
Using the product form of the auxiliary SPO measurement likelihood~\eqref{eq:A-SPO-measurement-likelihood}, 
\begin{align}
    & \mathrm{E}_{Z_k|X_k} \! \Big[ \! \log \! \tfrac{ p_{ \mathrm{A}\text{-}\mathrm{SPO}_k } ( Z_k | X_k,\, \paramPD^{\text{g}} ) }{ p_{ \mathrm{A}\text{-}\mathrm{SPO}_k } ( Z_k | X_k,\, \paramPD ) } \! \Big] \! = \cancel{ \mathrm{E}_{Z_k|X_k} \! \Big[ \log \tfrac{ p_{\ClutterRFS_k}(Z^0) }{ p_{\ClutterRFS_k}(Z^0) } \Big] } + \notag\\
    & \hspace{1cm} \textstyle \mathrm{E}_{Z_k|X_k} \! \Big[ \sum_{\mathbf{x}_k \in X_k} \! \log \! \tfrac{
    p_{\ObjectGenRFS(\mathbf{x}_k)} ( Z^{\mathcal{L}(\mathbf{x}_k)} | \mathbf{x}_k, \paramPD^g )
    }{
    p_{\ObjectGenRFS(\mathbf{x}_k)} ( Z^{\mathcal{L}(\mathbf{x}_k)} | \mathbf{x}_k, \paramPD )
    } \! \Big] \, .
\end{align}
The $D_{KL} \big( p \, \| \, q_\paramPD \big)$ in~\eqref{app:eq:KLD-mini-aux} thus can be substituted with
\begin{align}
    \! \Delta {=} \mathrm{E}_{X_k} \! \scalebox{1.3}{\Big[} 
    \textstyle \! \sum_{\mathbf{x}_k \in X_k} \! \mathrm{E}_{Z_k|X_k} \! \Big[ \log \! \tfrac{
    p_{\ObjectGenRFS(\mathbf{x}_k)} ( Z^{\mathcal{L}(\mathbf{x}_k)} | \mathbf{x}_k, \paramPD^g )
    }{
    p_{\ObjectGenRFS(\mathbf{x}_k)} ( Z^{\mathcal{L}(\mathbf{x}_k)} | \mathbf{x}_k, \paramPD )
    } \! \Big]
    \scalebox{1.3}{\Big]} \! . \! \label{eq:app:KLD-mini-adjusted}
\end{align}
Note that the measurements $Z^{\mathcal{L}(\mathbf{x}_k)}$ given $X_k$ are Bernoulli distributed.
Using~\cite[Eq.~(3.53)]{Mahler-Book:2014}, the inner expectation 
\begin{align}
    & \mathrm{E}_{Z_k|X_k} \! \Big[ \log \! \tfrac{
    p_{\ObjectGenRFS(\mathbf{x}_k)} ( Z^{\mathcal{L}(\mathbf{x}_k)} | \mathbf{x}_k, \paramPD^g )
    }{
    p_{\ObjectGenRFS(\mathbf{x}_k)} ( Z^{\mathcal{L}(\mathbf{x}_k)} | \mathbf{x}_k, \paramPD )
    } \! \Big] = \label{app:eq:inner-expectation-auxx}\\
    & \hspace{0.5cm} 
    \big( 1-P_{\ObjectGenRFS} (\mathbf{x}_k, X_k {\setminus} \{\mathbf{x}_k\}) \big) \log \Big( \tfrac{ 1-\paramPD^{g}(\mathbf{x}_k) }{ 1-\paramPD(\mathbf{x}_k) } \Big)
     + \notag \\
    & \hspace{0.5cm} \textstyle 
        P_{\ObjectGenRFS} (\mathbf{x}_k, X_k {\setminus} \{\mathbf{x}_k\}) \int \log \Big( \tfrac{ \paramPD^{g}(\mathbf{x}_k) \cancel{\likelihood( z | \mathbf{x}_k )} }{ \paramPD(\mathbf{x}_k) \cancel{\likelihood( z | \mathbf{x}_k )} } \Big) \likelihood( z | \mathbf{x}_k ) \d z 
    \,. \notag
\end{align}
Using Lemma~\ref{lemma-Palm-expectation}, the shorthand variable $\Delta$~\eqref{eq:app:KLD-mini-adjusted} becomes
\begin{align}
    & \hspace{-0.2cm} \Delta = 
    \textstyle \int \mathrm{E}_{O_k|\mathbf{x}_k} \! \big[ P_{\ObjectGenRFS} (\mathbf{x}_k, O_k)  \big] \log \Big( \tfrac{ \paramPD^{g}(\mathbf{x}_k) }{ \paramPD(\mathbf{x}_k) } \Big) \times 
    \label{app:eq:KLD-aux-sum} \\[-0.1cm]
    & \hspace{3cm} \textstyle \cancel{ \int \likelihood( z | \mathbf{x}_k ) \d z } \cdot D_{\RFSXi}(\mathbf{x}_k) \d \mathbf{x}_k \ + \notag\\
    & \textstyle \int \mathrm{E}_{O_k|\mathbf{x}_k} \! \big[ 1{-}P_{\ObjectGenRFS} (\mathbf{x}_k, O_k)  \big] \log \Big( \tfrac{ 1-\paramPD^{g}(\mathbf{x}_k) }{ 1-\paramPD(\mathbf{x}_k) } \Big) D_{\RFSXi}(\mathbf{x}_k) \d \mathbf{x}_k \, . \notag 
\end{align}
Since $\paramPD^g(\mathbf{x}_k) = \mathrm{E}_{O_k|\mathbf{x}_k} [ \, P_\ObjectGenRFS(\mathbf{x}_k, O_k) \, ]$, it follows that~\eqref{app:eq:KLD-aux-sum} is zero if and only if $\paramPD = \paramPD^g$, and thus $\paramPD^* = \paramPD^g$. 
\hfill $\square$

\begin{corollary}[Special case of Proposition~\ref{prop:best-fitting-KLD}]\label{corollary-special-case-of-proposition}
    Assume that $\paramPD$ is a function of only the mark in~\eqref{app:eq:KLD-mini-aux}, i.e., $\paramPD: \spaceOfMarks {\rightarrow} [0, 1]$.
    Take $\paramPD^g(m)$ to be $\mathrm{E}_{O_k,x_k|(\,\cdot\,,m)} [ \, P_\ObjectGenRFS\big((x_k,m)), O_k\big) \, ]$.
    Using Corollary~\ref{corollary-Campbell-Mecke-Marked}, the variable $\Delta$~\eqref{eq:app:KLD-mini-adjusted} can be adjusted further as
    \begin{align}
        \!\!\! & \hspace{-0.0cm} \! \Delta = \\[-0.1cm]
        & \textstyle \! \sum_{m\in \spaceOfMarks} \! \mathrm{E}_{O_k,x_k\!|(\cdot,m)} \! \big[ 1\!{-}\!P_{\ObjectGenRFS} \!\big( \!(x_k,m), O_k \!\big)  \big] \! \log \!\!\Big( \!\! \tfrac{ 1{-}\paramPD^{g}(m) }{ 1 {-} \paramPD(m) } \!\! \Big) \! D(m) \notag \! \\
        & \hspace{0cm}+ \textstyle \sum_{m\in \spaceOfMarks} \! \mathrm{E}_{O_k,x_k\!|(\cdot,m)} \! \big[ P_{\ObjectGenRFS} \!\big( \!(x_k,m), O_k\!\big)  \big] \! \log \!\!\Big( \!\! \tfrac{ \paramPD^{g}(m) }{ \paramPD(m) } \!\! \Big) \! D(m) . \! \notag
    \end{align}
    which is zero if and only if $\paramPD(m) $ is equal to $ \mathrm{E}_{O_k,x_k|(\,\cdot\,,m)} [ \, P_\ObjectGenRFS\big((x_k,m)), O_k\big) \, ]$, i.e, $\paramPD^* = \paramPD^g$.
    \hfill $\square$
\end{corollary}

\section{Proof of Corollary~\ref{corollary:objects-occlude-individually}}\label{app:proof-of-corollary:objects-occlude-individually}
Corollary~\ref{corollary:objects-occlude-individually} is a special case of the following lemma.

\begin{lemma}\label{lemma:aux-for-corollary-objects-occlude-individually}
    Let $p_{\mathbf{x}}$ and $p_{\mathbf{o}}$ be given continuous probability densities on a \emph{directionally limited}~\cite[p.~7]{Heinonen:AnalysisMetricSpaces:2001} (complete and) separable metric space $\X$, then the implication
    \begin{subequations}
    \begin{align}
        & \textstyle \int_{A_1} \! \int_{A_2} P_{\ObjectGenRFS}(\mathbf{x}, \!\{\mathbf{o}\}) \, p_{\mathbf{x}}(\mathbf{x})p_{\mathbf{o}}(\mathbf{o}) \, \d\mathbf{x}\d\mathbf{o} \notag\\[-0.1cm]
        & \hspace{1cm} \textstyle = \int_{A_1} \! \int_{A_2} P_{\ObjectGenRFS}(\mathbf{x}, \emptyset) \, p_{\mathbf{x}}(\mathbf{x})p_{\mathbf{o}}(\mathbf{o}) \, \d\mathbf{x}\d\mathbf{o} \label{app:eq:lemma-aux-Assumption} \\
        \Rightarrow \ & \textstyle \int_{B_1} \! \int_{B_2} P_{\ObjectGenRFS}(\mathbf{x}, \!O {\uplus} \{\mathbf{o}\}) \, p_{\mathbf{x}}(\mathbf{x})p_{\mathbf{o}}(\mathbf{o}) \, \d\mathbf{x}\d\mathbf{o} \notag\\[-0.1cm]
        & \hspace{1cm} \textstyle = \int_{B_1} \! \int_{B_2} P_{\ObjectGenRFS}(\mathbf{x}, \!O) \, p_{\mathbf{x}}(\mathbf{x})p_{\mathbf{o}}(\mathbf{o}) \, \label{app:eq:lemma-aux-Conclusion}\d\mathbf{x}\d\mathbf{o} \, ,
    \end{align}
    \end{subequations}
    holds for all measurable $A_1,A_2,\,B_1,B_2 \subseteq \X$.\\[0.2cm]
    \textsc{Proof:}
    The integrand in~\eqref{app:eq:lemma-aux-Assumption} is a product of integrable functions by Axiom~\ref{axiom:objects-occlude-individually} and thus locally integrable.
    Since $A_1$ and $A_2$ are arbitrary measurable sets on $\X$, open balls can be taken in the Lebesgue differentiation theorem\footnote{
        We use~\cite[Theorem~1.8]{Heinonen:AnalysisMetricSpaces:2001} under~\cite[Remark~1.13]{Heinonen:AnalysisMetricSpaces:2001} and~\cite[Example~1.15(f)]{Heinonen:AnalysisMetricSpaces:2001}.
        For $\X {=} \mathbb{R}^n$, no further assumptions would be needed assuming the reference measure is Radon~\cite[Remark~1.13]{Heinonen:AnalysisMetricSpaces:2001}.
    }~\cite[pp.~3-7]{Heinonen:AnalysisMetricSpaces:2001} to claim that the integrands in~\eqref{app:eq:lemma-aux-Assumption} are equal almost everywhere in $\X {\times} \X$, i.e., for all Lebesgue points
    \begin{align}
        P_{\ObjectGenRFS}(\mathbf{x}, \!\{\mathbf{o}\}) \, p_{\mathbf{x}}(\mathbf{x})p_{\mathbf{o}}(\mathbf{o}) = P_{\ObjectGenRFS}(\mathbf{x}, \emptyset) \, p_{\mathbf{x}}(\mathbf{x})p_{\mathbf{o}}(\mathbf{o}) \, .
    \end{align}
    Therefore, wherever $p_{\mathbf{x}}(\mathbf{x}) {\neq} 0$ and $p_{\mathbf{o}}(\mathbf{o}) {\neq} 0$, i.e., for almost all $\mathbf{x} {\in} \mathrm{supp}(p_{\mathbf{x}})$ and $\mathbf{o} {\in} \mathrm{supp}(p_{\mathbf{o}})$, we have also that $P_{\ObjectGenRFS}(\mathbf{x}, \!\{\mathbf{o}\}) {=} P_{\ObjectGenRFS}(\mathbf{x}, \emptyset)$, with $\mathrm{supp}$ being the support. Thus by Axiom~\ref{axiom:objects-occlude-individually}, we also have that $P_{\ObjectGenRFS}(\mathbf{x}, O {\uplus} \{\mathbf{o}\}) {=} P_{\ObjectGenRFS}(\mathbf{x}, \!O)$ for almost all $\mathbf{x} {\in} \mathrm{supp}(p_{\mathbf{x}})$ and $\mathbf{o} {\in} \mathrm{supp}(p_{\mathbf{o}})$.
    Since $p_{\mathbf{x}}(\mathbf{x})$ and $p_{\mathbf{o}}(\mathbf{o})$ are zero almost everywhere in the complements of $\mathrm{supp}(p_{\mathbf{x}})$ and $\mathrm{supp}(p_{\mathbf{o}})$, respectively, we have that 
    \begin{align}
        P_{\ObjectGenRFS}(\mathbf{x}, \!O {\uplus} \{\mathbf{o}\}) \, p_{\mathbf{x}}(\mathbf{x})p_{\mathbf{o}} = P_{\ObjectGenRFS}(\mathbf{x}, \!O) \, p_{\mathbf{x}}(\mathbf{x})p_{\mathbf{o}}(\mathbf{o}) \, , \label{app:eq:lemma-auxiliary-result}
    \end{align}
    almost everywhere in $\X$.
    Integrating~\eqref{app:eq:lemma-auxiliary-result} on the prescribed $B_1$ and $B_2$ yields~\eqref{app:eq:lemma-aux-Conclusion} which concludes the proof. \hfill $\square$ 
\end{lemma}

Corollary~\ref{corollary:objects-occlude-individually} results directly from Lemma~\ref{lemma:aux-for-corollary-objects-occlude-individually} by setting $A_1 {=} A_2 {=} B_1 {=} B_2 {=} \X$ and noting that $p_{\mathbf{o}}(\mathbf{o})$ integrates to one on $\X$.
Note that if the densities are, e.g., Gaussian, the assumption~\eqref{app:eq:lemma-aux-Assumption} can be satisfied only approximately.

\bibliographystyle{ieeetr} 
\bibliography{references} 

\end{document}